\documentclass[acmsmall]{acmart}
\usepackage{soul}
\usepackage{subcaption}
\usepackage{graphicx}
\usepackage[table,xcdraw]{xcolor}



\begin{document}

\title[Fostering Reflections on Designer Inclusion in OSS Through Speculative Design]{What If We Work Together? Fostering Reflections on Designer Inclusion in Open Source Software Through Speculative Design}

\author{Rozhan Hozhabri Nezhad}
\orcid{0009-0009-8741-5295}
\email{rozhan.hozhabri-nezhad@polymtl.ca}
\affiliation{%
 \institution{Polytechnique Montreal}
 \city{Montreal}
 \state{Quebec}
 \country{Canada}}

\author{Jin L.C. Guo}
\orcid{0000-0003-1782-1545}
\email{jguo@cs.mcgill.ca}
\affiliation{%
 \institution{McGill University}
 \city{Montreal}
 \state{Quebec}
 \country{Canada}}

\author{Jinghui Cheng}
\orcid{0000-0002-8474-5290}
\email{jinghui.cheng@polymtl.ca}
\affiliation{%
 \institution{Polytechnique Montreal}
 \city{Montreal}
 \state{Quebec}
 \country{Canada}}

\begin{abstract}
    Open source software (OSS) often prioritizes technical functionality over usability and UX design. This imbalance limits OSS adoption among broader, non-technical users. Key underlying factors contributing to this issue are the shortage of design expertise in OSS and a dominant developer-centric mindset. To address these persistent issues, we explore the potential of speculative design as a catalyst for transforming the OSS community's mindset towards a more designer-inclusive environment. Our design was informed by an analysis of online forums, which revealed designers' motivations and challenges when contributing to OSS. Guided by these insights, we created two speculative societies, Husia (collectivist) and Reetar (individualist), in which designers are valued for different reasons and their work incorporated in different ways. Through a user study with 12 OSS practitioners (seven designers and five developers), we found that our speculative societies provoked participants' rich and critical reflections on OSS values, the root causes of challenges, and proposed actions. Our work provides insights into how speculative design can be used in the practical, sociotechnical context of OSS to stimulate critical reflection, improve awareness, and yield recommendations for fostering an equitable, sustainable, and inclusive OSS environment. 
\end{abstract}

\maketitle
\section{Introduction}
Open source software (OSS) development as a collaborative, community-driven activity has experienced significant growth over the past decades, playing a major role in the production of widely used software solutions~\cite{Brian2006}. Despite its broad adoption, UI/UX design in OSS often receives comparatively little attention~\cite{sanei}. These important software aspects typically play a secondary role to technical functionality, which can result in software that is powerful yet cumbersome for non-technical users~\cite{andreasen, Raza2012}. Consequently, many OSS products failed to cater to a broad spectrum of users, particularly those lacking technical expertise~\cite{eric,rolandssona}.

One key reason for this imbalance in emphasis between UI/UX design and technical functionalities, as Nichols and Twidale~\cite{nichols2003} pointed out and echoed in many other studies~\cite{powerempower, bach}, is that ``\textit{usability experts do not get involved in OSS projects.}'' The underlying causes of this long-lasting problem are manifold and can be traced back to the persistent developer-led and development-driven mindset within OSS communities~\cite{feller_perspectives_2005, wang, hellman_facilitating_2021}. Tools for OSS collaborative creation, such as version control systems and issue trackers, are developer-centric and perplexing to non-technical contributors such as designers and end-users~\cite{hellman_facilitating_2021}.  When it comes to designer-developer collaboration, the common challenges, such as separated decision-making procedures, different professional behaviors, and a lack of mutual understanding~\cite{Zhang2025WhoBlame}, are magnified through the stronghold of the developer-centric mindset of OSS.

As a result, improving the UI/UX design of OSS requires more than just tweaking existing processes or tools. It needs to start from a fundamental shift in the mindset of the OSS community, although this change will likely be slow and gradual. We believe that speculative design~\cite{Dunne2013,Auger} can catalyze this shift by imagining alternative futures that challenge existing norms and provoke critical reflection. While speculative design has been shown to foster discussion around complex and value-laden issues in other domains, such as contestability~\cite{alfrink} and sustainability~\cite{Chopra2022}, it has not yet been explored in the context of OSS, particularly regarding how stakeholders negotiate conflicting values and build social infrastructure.

Motivated by this perspective, and using a Research-through-Design approach~\cite{Gaver2012RtD,Zimmerman2007RtD}, we created two speculative societies and imagined how collaborative software development would be performed in each. Both societies value open collaboration and design contributions but differ in their underlying values---one collectivist and one individualist, informed by Hofstede's cultural model~\cite{hofstede}, prior OSS literature~\cite{hars,Jahn2025,Gerosa2021}, and our analysis of forum discussions on OSS designers' motivations and challenges. These contrasting scenarios were designed to amplify value tensions in the existing OSS practice while evoking participants' own experiences. We presented textual descriptions and visual storyboards of these societies in a user study with OSS contributors. With the stimulation provided by these speculative scenarios, participants' critically reflected on their fundamental values, the barriers hindering OSS design incorporation, and potential actions to improve design integration. Participants provided rich insights into how OSS communities might show greater appreciation for design work, foster inclusive environments for both new and experienced designers, and improve tooling to better incorporate design in the OSS workflow.

The main contributions of this paper are twofold. First, through the user study to gather OSS contributors' rich and critical reflections provoked by the speculative societies, we surface specific barriers and concrete enablers for designer inclusion in OSS workflows. We offer actionable guidelines, grounded in participants' feedback, for integrating tools and processes into existing OSS communities to support designer participation. Second, we reflect on how speculative design can be adapted to the domain of supporting designer involvement in OSS. By examining our own design actions and participants' responses, we gain valuable insights that can inform similar future efforts to apply the speculative design approach. Together, these contributions provide useful information and a practical road map for fostering designer-inclusive OSS projects.

\textbf{Researcher Positionality.}
We acknowledge that our positionality as researchers shaped how we approached speculative design and how we collected, analyzed, and interpreted the data. Our expertise and background are rooted in the intersection between human-computer interaction (HCI) and software engineering (SE). Two of the authors have had prior education and varied experience in interaction design. Two authors also had extensive prior experience engaging with OSS communities and studying the aspects of usability, design, and inclusiveness in the OSS context. All authors are grounded in a qualitative, interpretivist stance, with an emphasis on empirical methods. These backgrounds, expertise, and beliefs influenced our research direction and approach.

\section{Background and Related Work}
In this section, we briefly review related work on (1) the OSS development process and its UI/UX design and (2) speculative design and its potential to provoke reflections on OSS design.

\subsection{OSS and Its UI/UX Design}
\label{sec:lit_OSSDesign}
OSS has significantly reshaped the software development landscape, fostering collaborative and community-driven innovation~\cite{chelkowski, gatekeeping}. The OSS development process is often decentralized, drawing on contributions from a community of developers, maintainers, and contributors~\cite{unveil, lessonsoss, jahn}. Unlike traditional software development, which follows structured workflows and predefined roles, OSS projects can operate through a more fluid and self-organized model, where decisions emerge through discussions~\cite{Arya_InfoType_19, Gilmer_summit_23}, consensus-building~\cite{bogart,unveil}, and sometimes conflict~\cite{Ferreira2021}. While this model enables rapid iteration and stimulates innovation, it can also lead to inconsistencies in project priorities, particularly in areas like UI/UX design, which often receive less attention compared to core technical functionalities~\cite{sanei, wang}.

As a result, usability and UI/UX design have long been a critical challenge for OSS. Historically, OSS projects have prioritized technical functionality over user experience, often neglecting the needs of novice and non-expert users~\cite{khalajzadeh, bach, dawood}. Design concerns in OSS are further compounded by limited resources, lack of dedicated design expertise, and cultural barriers within OSS communities~\cite{andreasen, wang, nichols, llerena25}. To address these challenges, various strategies have been proposed to integrate design-related practices into OSS development without compromising its collaborative nature. For example, participatory methods, which actively involve users and other stakeholders in the design process, have been proposed as a potential solution to bridge the gap between technical development and user needs~\cite{partdesign}. Within the participatory framework, various hybrid governance models, incorporating structured collaboration between commercial entities and OSS communities, were found to affect the openness and the collaboration dynamics of the communities~\cite{maenpaa}. Efforts have also been made through integrating UI/UX practices early in OSS development and fostering partnerships between developers and design professionals, aligned with user-centered design principles, to improve the acceptance and impact of design contributions~\cite{rajanen, ossdev, llerena25}. 

Nevertheless, the proposed solutions to improve usability and UI/UX in OSS face persistent obstacles and limitations. Strategies like fostering a design-centered culture and deeper user engagement are often too broad, resource-intensive, or dependent on the willingness of developers to prioritize design~\cite{raza, wang, geiger}. Feedback mechanisms may disproportionately reflect vocal minority views, while participatory methods can be difficult to scale and sustain due to knowledge loss and contributor burnout~\cite{chengguo, Wang2020ArguLens, Ferreira2021}. Moreover, the philosophical commitment of OSS to transparency and collaboration sometimes clashes with the structured practices required for effective design work~\cite{powerempower, opensourcehci, dawood}. These tensions highlighted the importance of reconciling OSS values with user-centered design practices, potentially through inducing voluntary reflections on the OSS status quo among the current community members. This is an issue targeted by our research.

\subsection{Speculative Design}
Speculative design is an approach that focuses on exploring thought-provoking designs to reflect on the hidden status quo and explore future possibilities. Instead of creating design solutions for current problems, it often aims at challenging existing assumptions, questioning societal norms, and stimulating critical discourse about the implications of emerging technologies and possible future scenarios~\cite{Dunne2013, Auger}. The speculative design approach has found widespread application in various societal and technological contexts, such as wearable tracking in sports~\citep{kolovson}, novel sensory experiences~\citep{besevli}, and the future of biodata usage and engagement~\citep{tsaknaki}, to name a few. 

Although there is no direct study on speculative design in the context of OSS, the speculative design approach can play a transformative role in addressing persistent challenges through a creative lens. For example, speculative scenarios might envision alternative governance models or tools that support collaboration between designers and other involved parties~\cite{farias, zhang}. Adopting such approaches in OSS could enable OSS stakeholders to critically examine existing limitations and envision more inclusive and productive systems. This emphasis on creating reflective opportunities aligns with the broader principle of ``reflection in action'' of design~\cite{schon}. In this context, \citet{sengers} proposed that critical reflection (i.e., making unconscious aspects of experience consciously available for choice) should be a core element of technology design. They argue that reflection is often triggered by an element of surprise, which disrupts routine patterns of knowing-in-action and compels individuals to reconsider their assumptions and practices to challenge the status quo and stimulate critical inquiry. \citet{bell} expanded on the concept of defamiliarization, a process of making the familiar strange, to prompt critical reflection on cultural and technological norms. By exposing participants to unexpected scenarios or reframing everyday objects, defamiliarization invites them to question taken-for-granted assumptions and explore new possibilities.

By creating imaginative scenarios and alternative futures, speculative design has the potential to challenge participants' assumptions and facilitate innovative solutions. \citet{Auger} emphasizes the importance of the ``perceptual bridge,'' wherein speculative artifacts connect abstract concepts with tangible representations. \citet{sondergaard} explored ``fabulation'' as a speculative practice for design futuring, emphasizing the importance of storytelling. This narrative-driven approach aligns with speculative design's emphasis on engaging participants both intellectually and emotionally, and it inspired our approach of creating narratives and storyboards in our speculation process. Similarly, \citet{kozubaev} identified five reflective modes that speculative design enables. Among these modes, our work aligns closely with the \textit{Engaging with the Real World} mode, as we utilize speculative design to interrogate the social structures and existing dynamics of designer involvement in OSS. We further address the \textit{Positionality} mode by centering a marginalized group (i.e., designers) that reflects our own identities, as well as the \textit{Knowledge Production} mode by surfacing insights towards deeper structural challenges and potential pathways for integrating design into OSS. Overall, these previous works highlighted characteristics that make speculative design particularly effective in complex sociotechnical domains, such as OSS, where structural and cultural barriers often hinder collaboration. We take on this perspective and explore how outcomes of this approach could encourage OSS community members to engage deeply with the persistent challenges and possible solutions regarding UI/UX design in OSS.

\section{Preliminary Investigation to Inform Speculative Design}
To inform our speculative design process, we first sought to understand what motivated designers to engage with OSS projects and the challenges they encountered.

\subsection{Methods}
To achieve our goal, we focused on collecting design-related topics discussed in two online open-source forums: the r/opensource\footnote{https://www.reddit.com/r/opensource/} subreddit and the Open Source Design\footnote{https://discourse.opensourcedesign.net} forum. Data collection happened in November 2023. These two forums were selected as data sources because they have a substantial number of members with diverse backgrounds (both designers and developers), are active forums with daily user activity, and are dedicated to a wide range of open source topics. For the r/opensource, we searched for the keyword ``\textit{design}'' on the entire subreddit to capture design-related posts. For the Open Source Design forum, we collected all posts in the ``design lounge'' where community members discussed design-related topics. This process yielded 248 posts on Reddit and 211 posts on the Open Source Design Lounge.

We then familiarized ourselves with the collected data and filtered the posts according to their relevance to the topic of designer engagement. Particularly, we included posts touching on (1) design needs in OSS, (2) contributions of design work to OSS, (3) the UI/UX design practice of specific OSS products, and (4) requests for or provision of resources and expertise in UI/UX design. After an iterative filtering process, we identified 34 relevant posts from the r/opensource subreddit and 31 relevant posts from the Open Source Design forum. Within this sample, the posts from the r/opensource subreddit had a median of 7 comments ($IQR=11$, $total=392$) and involved a median of 5 unique participants ($IQR=8$), whereas the posts from the Open Source Design forum had a median of 5 comments ($IQR=9$, $total=260$) and involved a median of 4 unique participants ($IQR=3$). Thus, in total, we analyzed 65 posts and 652 comments across the two forums. 

The subsequent data analysis process was qualitative and inductive, following recommendations of \citet{Bingham2023Qualitative}. It involved multiple rounds of iterative coding, synthesizing, and theme identification to examine the distinct aspects of designers' motivations and challenges to engage with OSS. All three authors were involved in the data analysis process. The first author led the analysis by identifying the initial sets of codes and themes. Then all authors iterated the coding through multiple sessions of discussion, individual reflection, and consolidation of opinions. The initial rounds of coding were conducted using Atlas.ti. The resulting codes and quotes were then exported to Miro, where we conducted a collaborative affinity diagramming exercise for further grouping, theme development, reflection, and interpretation. The data analysis terminated when prominent and cohesive themes were extracted and all authors were confident that the themes were rich enough to inform the speculative design.

\subsubsection*{Considerations and Limitations}
We performed our preliminary study by qualitatively analyzing the OSS forums, considering their roles in capturing the daily communication of the community members, documenting their questions, concerns, and responses in an unobtrusive manner~\cite{hineVirtualEthnographyModes2008}. While methods such as interviews and surveys can also prompt participants to reflect on their experience of OSS contributors, they do not afford the visibility into the naturally occurring, spontaneous, and peer-to-peer interactions in online communities where practices, challenges, and norms unfold in situ. Furthermore, previous work (such as \cite{bach, wang}) has adopted methods such as interviews and surveys to understand designers' and developers' concerns in OSS. We use their output to inform our analysis of the forums, instead of reinventing the wheel.

At the same time, we acknowledge that the social dynamics of the general OSS forums, such as the r/opensource subreddit, might mirror the existing dynamics of OSS and therefore create barriers for designers to participate. To overcome this challenge, we also selected the other designer-oriented forum, Open Source Design, in our analysis. Still, our findings can be limited by the context allowed to be communicated in these forums, and other factors in real projects might be overlooked.

\subsection{Results}
% Our analysis uncovered several key insights related to designers' motivations and challenges in participating in OSS.

\subsubsection{Designers' motivations to contribute to OSS projects}
\label{subsubsec:designer_motivations}
Our aim in grasping the motivations of designers was to explore how we can meet these needs and desires in later speculative scenarios.

\textbf{Enhancing OSS usability and user experience:}
The designers on the analyzed forums often cited their motivation to engage in OSS as their concern for its often inferior usability and user experience. For example, a designer mentioned: ``\textit{A common complaint about open source software is the lack of focus on usability and design,}'' and explained that this motivated them to figure out how to contribute to OSS. Another designer also mentioned, as a reason for getting into OSS, ``\textit{I want to see a greater focus on user experience and interface design.}''

\textbf{Personal growth:} 
Designers were also motivated by personal growth and career opportunities, particularly improving their portfolios and enhancing job prospects. For instance, a designer who was seeking an open source project to contribute mentioned: ``\textit{At the moment I have a little time to spare, and as preparation for a potential job (which could change my life for 100000x better), I would love to get some experience doing UX design!}'' 
% Another designer stated: ``\textit{I am looking for some projects to work on for my portfolio.}''

\textbf{Giving back to the community:} 
Some designers mentioned that they were motivated by a sense of social responsibility to give back to the OSS community. For instance, a designer pointed out, ``\textit{[Open source software] helped me a lot getting into Graphic Design --- all the tools I use (including OS) are Open Source --- and I would like to return the favor while building my own set of skills.}'' 

\subsubsection{Challenges designers faced when contributing to OSS}
\label{subsubsec:designer_challenges}
We identified the following main challenges discussed by the forum users. Understanding these challenges would help us identify designers' pain points and potentially find design opportunities to address them.

\textbf{Conflicts between designers' values and the philosophy of OSS:}  In the forum discussions, we found some designers indicated that the open philosophy embraced by OSS somewhat contradicts the designers' values regarding their own profession and work. For example, a designer mentioned: ``\textit{Designers have struggled hard to be recognized as a real profession, so when you come to open source, suddenly you're giving away your work for free.}'' Moreover, copyright and ownership of design work were also a major concern, as a designer cited a ``\textit{fear of copyright `loss'}'' as a barrier for designers to adopt OSS collaboration methods.

\textbf{Designers' lack of knowledge about OSS:}
Related to the previous theme, many forum members mentioned that designers were unfamiliar with the concept, value, and development process of OSS. As a community member summarized: ``\textit{Designers don't know what OSS is, how it's developed, why it's important, and what impact it has on technology and on society. In short, they are not familiar with the politics and ideologies of OSS.}'' Some attributed this gap to the education system of the profession, as a designer noted: ``\textit{Most design schools/universities do not teach, nor are educated on open source.}'' This lack of knowledge often led to misconceptions and confusion about how designers can meaningfully contribute to OSS.

\textbf{Lack of support for designer-developer collaboration in OSS:}
Both designers and developers noted the difficulty in finding collaborators from the other profession in the OSS context. Many mentioned the need for a dedicated platform to improve communication and collaboration. For example, a designer ideated a hub where ``\textit{Developers could submit requests about visual needs (icons, logo, visual identity), specifying budget, deadlines, purpose, etc. Designers can submit artwork on the platform so that other people can see, comment, and even contribute to the visual.}'' 

\textbf{Devaluation of designers and their work in OSS:}
Another common challenge mentioned by designers is that they frequently felt themselves and their work undervalued by the OSS communities. For instance, a designer explained: ``\textit{[Developers] don't see the usefulness in bringing a UX designer or graphic designer into the fold of the project, or worse still, think that they can do the job of UX design because `how hard can it be anyway?'}'' This divide between developers and designers was often attributed to OSS project structures, limited resources, and misalignment between the typical monolithic design processes and OSS's incremental workflows. For example, related to the last aspect, a designer wrote: ``\textit{Design is hard to split up in parts which you patch together.}''

\section{Speculative Design Process and Outcomes}
Inspired by the results of the forum analysis, we conducted speculative design to create fictional societies and systems. Our goal was to explore alternative ways OSS could function, particularly in relation to design and the role of designers. We intended to use these fictional societies to introduce unique elements into our discussions with OSS practitioners, encouraging reflection on the status quo of OSS UX design, its limitations, and possible solutions.

\subsection{Design Rationales and Design Process}
OSS development is often considered a community-driven activity, where community members contribute by sharing ideas, knowledge, and resources, as well as creating artifacts to collectively advance projects. Through our forum analysis, we found that designers were motivated to contribute to OSS for both collective reasons (i.e., enhancing OSS design and giving back to the community) and personal interests (i.e., for personal growth). These factors echo with findings from previous studies that OSS developers were motivated by both internal factors (e.g., intrinsic motivation and altruism) and external rewards (e.g., future returns and personal needs)~\cite{hars,Jahn2025,Sun2025Chase,Gerosa2021}. It also aligns with \citet{vonkrogh}'s concept of the ``\textit{private-collective}'' innovation model of OSS, which describes how OSS community members pursue private benefits---such as reputation enhancement and career advancement---while simultaneously contributing to an effort that benefits the wider community. Prior work on the impact of cultural values on developer behavior (e.g., the work by \citet{zold} about how individualist and collectivist cultural values affected developers' participation in online communities) further indicated the importance of the individualism-collectivism cultural dimension in the context of OSS~\cite{hofstede}.

Building on these insights, and through iterative analysis-design-reflection cycles~\cite{Zimmerman2007RtD}, we gradually converged our focus in speculative design to explore the tension between the individualist and collectivist interests in OSS. Particularly, we designed two extreme speculative societies: one, named \textbf{Husia}, entirely collectivist, and the other, \textbf{Reetar}, individualist. These two contrasting scenarios allowed us to put the audience in opposing situations drastically different from our current world, while reminding them of familiar experiences. Moreover, since our main objective was to explore how designers and their work can be respected, valued, and embedded in OSS, we ensured that both speculative societies prioritized designers' work as a crucial element.

We began by imagining scenarios and characteristics of the societies, incorporating the findings from our forum analysis related to the designers' motivations and challenges for participating in OSS projects. For example, our re-imagination of the design-centric mindset in both societies targeted the challenge of the undervaluation of designers in current OSS contexts. However, this mindset may be supported by different values, beliefs, and cultural norms in the two extreme societies; while people in the collectivist society (\textit{Husia}) might promote design to produce products for common benefits, those in the individualist society (\textit{Reetar}) might do it to maximize personal gain and minimize liability. Moreover, forum users mentioned that a prominent challenge was the lack of support for designers to collaborate with other designers and developers. Designers were also often overwhelmed during onboarding due to the lack of knowledge and experience about OSS. To address these challenges, we speculated on the platforms in both societies that support contributor onboarding and allow designers to work together and with other stakeholders. Such collaborative platforms, however, have different underlying driving forces in the two societal extremes. In \textit{Husia}, the value of togetherness pushes their adoption; whereas in \textit{Reetar}, collaboration efforts are fuelled by practical needs and self-preservation. Additionally, a crucial takeaway from our forum analysis is that the obstacles to integrating designers into OSS communities cannot solely be attributed to the developer-centric culture and practices within OSS. Rather, we found that designers also harbour mindset issues that can be at odds with the principles of openness inherent in OSS. To address this, our designs deliberately targeted both designers and developers.

A key consideration that guided our speculative exploration was finding an equilibrium between fiction and familiarity. The speculative scenarios should be provocative enough to spark reflection and discussion, yet remain relevant and meaningful to participants' real experiences. This is similar to \citet{bell}'s consideration in the context of smart home technologies for opening up the design space. In our study, the challenge was to create fictional societies where the OSS development process fits naturally. As we wanted to trigger reflections to focus mainly on designers' involvement and the design process itself, we needed to include elements that show a focus on design without simply copying the current industry best practices or ideal OSS development processes. During the speculation process, we rooted these elements in the insights collected from the forum analysis, which helped us ground our societies in relatable considerations. We then crafted extreme solutions to those problems to create a factor of surprise. For example, the forum users mentioned that smooth onboarding or clear guidelines are rare for designer involvement; so we created a ``Central Board'' in \textit{Husia} to practically vanquish these pain points by handholding designer onboarding. The forum users also discussed a general devaluation of designers' work in OSS, which we directly confronted in \textit{Reetar} by introducing a reputation system with extremely high risk to necessitate design work. This type of defamiliarization within familiar contexts provoked new perspectives from participants, allowing them to question the existing beliefs and assumptions.

Based on these considerations, we derived the core concepts of the two societies. For \textit{Husia}, the collectivist society, we pushed towards the sense of common benefits and the value of being together, and thus focused on a small and close society in which ownership is not prioritized and personal credit is irrelevant. For \textit{Reetar}, on the other hand, we introduced reputation points (RP) associated with each individual as a form of currency, gained by contributing to well-received products and used to get access to more promising projects, to consolidate the individualist values of this society and the necessity of openness and collaboration. Table~\ref{tab:mapping_design_forum} summarizes the main considerations of these two societies and how the motivations and challenges of OSS contribution derived from the forum analysis influenced those considerations.

For each society, we wrote an initial description to capture its core cultural values, social dynamics, and technological landscape. We refined the details of these societies over several iterations, informed by reflections and discussions among ourselves, as well as informal feedback from a few OSS contributors. 
Several changes were made during these iterations. For example, an initial version of the societies tended to evoke distinct moods---one suggesting an ideal society and the other a dystopian one. In response, we revised the descriptions by adopting neutral language, incorporating both positive and negative aspects, and removing unnecessary details. 

To allow our audience, both designers and developers, to relate to the narratives, we divided each society into two parts. The first part provided an overview of the society's characteristics and a short description of the software development process, while the second part offered a walkthrough of a person living in that society as either a designer or a developer. This second part eventually became storyboards that provided visual support of a dynamic narrative from an OSS contributor's perspective. The first author recorded a voiceover explaining the storyboard and integrated it into a video. Intentionally, we left some details open in these society descriptions and storyboards to encourage participants' imagination and provoke questions that would foster critical reflection.

\begin{table}[t]
\definecolor{shade}{RGB}{229, 229, 255}
\definecolor{shade2}{RGB}{255, 229, 229}
\centering
\caption{Main characteristics of the speculative societies, Husia and Reetar, informed by the forum analysis.}
\footnotesize
\begin{tabular}{
    >{\raggedright\arraybackslash}p{2.7cm}
    >{\raggedright\arraybackslash}p{3.4cm}
    >{\raggedright\arraybackslash}p{3.4cm}
    p{0.01cm}
    >{\raggedright\arraybackslash}p{2.6cm}}
\toprule
& \textbf{Husia} & \textbf{Reetar} & &\\
\midrule[0.05pt]

\cellcolor{shade}Designers' Motivations (see Sec.~\ref{subsubsec:designer_motivations}) & \cellcolor{shade}\textit{Enhancing OSS design}; \textit{Giving back to the community}  & \cellcolor{shade}\textit{Personal Growth}  & & \cellcolor{shade2}Designers' Challenges (see Sec.~\ref{subsubsec:designer_challenges}) \\

\cmidrule{1-3}

Value of Design & Produce products for common benefits & Maximize personal gain and minimize liability & & \cellcolor{shade2}\textit{Devaluation of designers and their work} 
\\&&&&\cellcolor{shade2}\\

Drive for Collaboration & The fundamental value of togetherness & Individual practical needs and self-preservation & & \cellcolor{shade2}\textit{Lack of support for designer-developer collaboration in OSS} 
\\&&&&\cellcolor{shade2}\\

Reason for Openness & Usefulness to more people in the society & Innovation and extensibility lead to individual gain & & \cellcolor{shade2}\textit{Conflicts between designers' values and the philosophy of OSS}
\\&&&&\cellcolor{shade2}\\

Collaborative Platforms & A `central board' and project-specific spaces that learns and automatically adjusts to each worker's style and needs while maximizing the expected benefit of the entire society & Virtual, smart spaces that connects developers, designers, and other experts to work together to earn Rupation Points (RP) & & \cellcolor{shade2}\textit{Designers' lack of knowledge about OSS}; \textit{Lack of support for designer-developer collaboration in OSS} \\
\midrule

Cultrual Tendency~\cite{hofstede} & \textbf{Collectivist} & \textbf{Individualist} & \\
\bottomrule

\end{tabular}
\label{tab:mapping_design_forum}
\end{table}

\subsection{Speculative Societies}
\begin{figure}[t!]
    \centering 
    \begin{subfigure}{0.45\textwidth}
    \includegraphics[width=\textwidth]{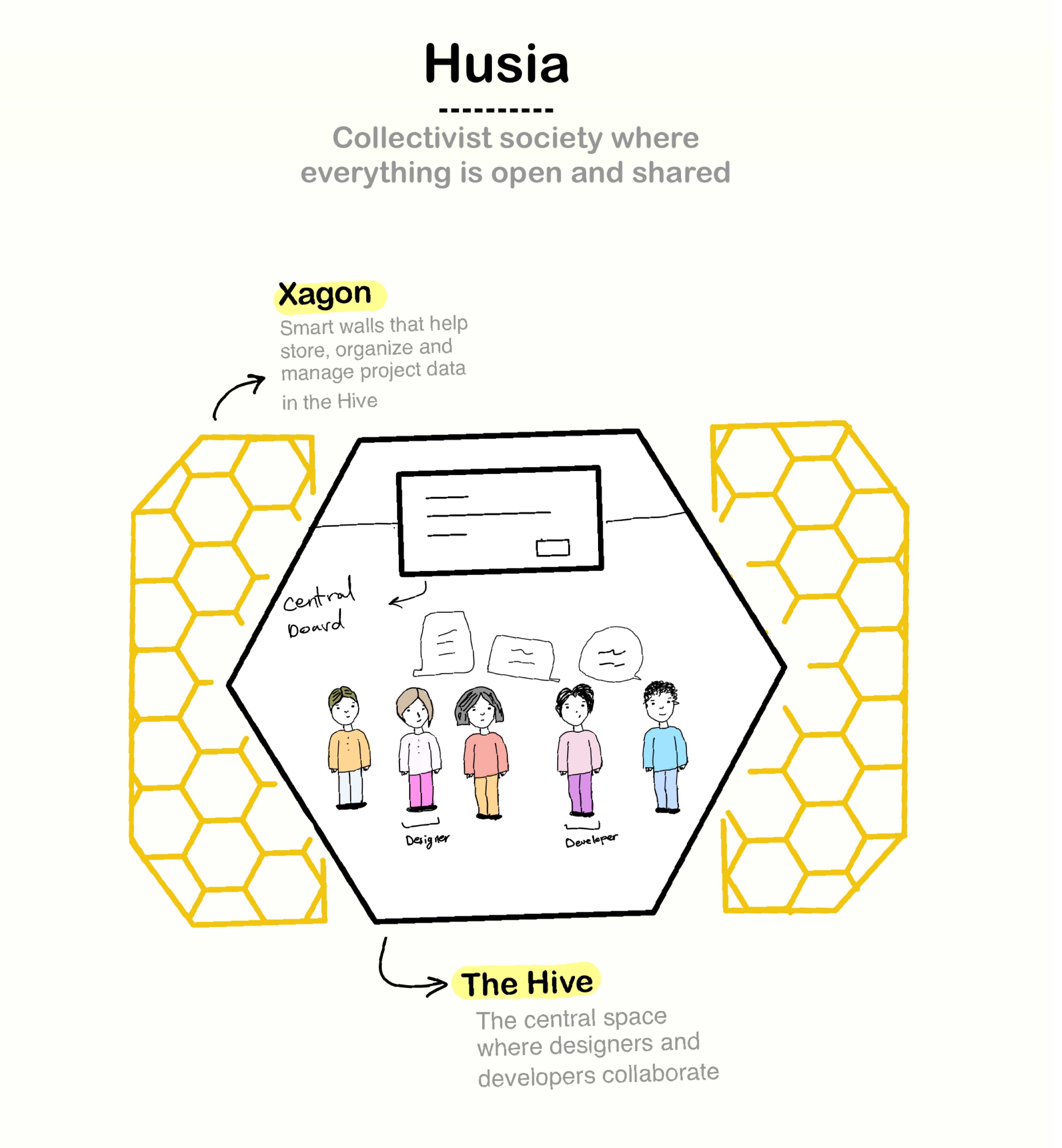}
    \caption{A visual summary of Husia}
    \label{fig:thumbnail_husia}
    \end{subfigure}
    \hfill
    \begin{subfigure}{0.45\textwidth}
    \includegraphics[width=\textwidth]{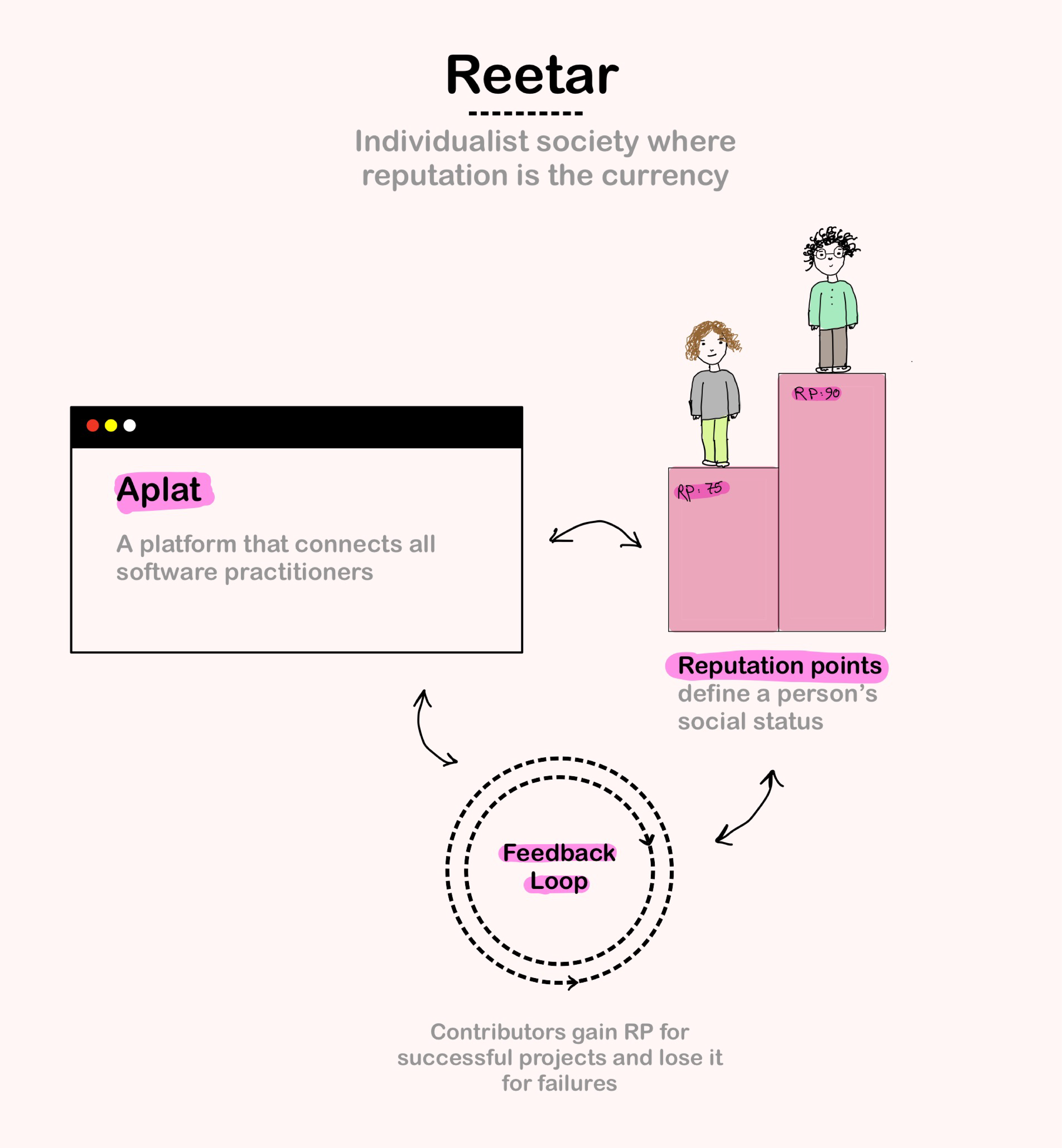}
    \caption{A visual summary of Reetar}
    \label{fig:thumbnail_reetar}
    \end{subfigure}
    \caption{Concept images of the two speculative societies: \textit{Husia} and \textit{Reetar}, one collectivist society and the other individualist society.}
    \label{fig:thumbnails}
\end{figure}

Below, we provide brief descriptions of the two speculative societies that we created, \textit{Husia} and \textit{Reetar}, along with summaries of the storyboards walking through a day in the life of a designer/developer in each world. Figure~\ref{fig:thumbnails} presents the main concepts of the two societies. The full descriptions and the storyboards used in the user studies are in the Appendix.

\subsubsection{Husia (see Figure~\ref{fig:thumbnail_husia} and Appendix~\ref{sec:husia_details})}
This is a small, closely connected society where everything is shared by all and personal ownership does not exist. Community, collaboration, and collective benefit guide all aspects of life, with no focus on individual credit or monopolization. Children are educated together and guided toward professions that match their strengths, always with the goal of contributing to society. All knowledge, products, and software are openly available for anyone to use, modify, and improve. This collaborative spirit is embodied in the Hive, a shared physical space where intelligent systems (Central Board and Xagons) support design and development, adapting to each worker's abilities while optimizing outcomes for the entire community. The full description of Husia presented to the participants is available in Appendix~\ref{sec:husia_description}.

The full storyboard of Husia is available in Appendix~\ref{sec:husia_storyboard}. It describes how Fiona, a recent design graduate, enters the Hive and is onboarded by the Central Board. As a first-time contributor, she reviews recommended projects on the Central Board, chooses a project aligned with her interests, and is matched with suitable UI and graphic design tasks. After accepting them, she is directed to the project's Xagon---a dedicated physical workspace equipped with smart walls that support co-located collaboration, preserve project history, and integrate user feedback. Within the Xagon, Fiona redesigns the landing page of the system, using reminders provided by the Xagon about user needs and existing design guidelines. Once she submits her draft, the Xagon then flags the design for developers to review, keeps them informed of the design's user impact, and finally marks the task as semi-completed to invite team-wide feedback.

\subsubsection{Reetar (see Figure~\ref{fig:thumbnail_reetar} and Appendix~\ref{sec:reetar_details})}
This is a society where reputation is the primary currency, measured through Reputation Points (RPs) that determine social status and opportunities. Individuals gain or lose RPs based on how others react to the products or services they provide, with high stakes since losses can be permanent. Innovation is essential for earning significant RP, encouraging both solo work for maximum reward and group collaboration to spread risk. Software development centers around Aplat, a shared platform where developers and designers collaborate, but success and failure directly affect reputation, making cooperation cautious and highly competitive. Figure~\ref{fig:thumbnail_reetar} presents the main concepts of Husia. Its full description presented to the participants is available in Appendix~\ref{sec:reetar_description}.

The full storyboard of Reetar is available in Appendix~\ref{sec:reetar_storyboard}. It describes how Alex, an aspiring developer, chooses to join a team project instead of freelancing in order to reduce risk and build RPs by working on a familiar project through Aplat. Aplat structures collaboration around the design process, requiring early design artifacts and providing tools to ensure design-driven decisions. Each project is guided by a design master who aligns user needs with RP priorities and manages risks. Throughout development, RP considerations shape every decision, while Aplat records goals, plans, and risks so teams remain aligned. After joining a designer-initiated task, Alex collaborates with a designer in Aplat, where real-time feedback enables close coordination. When feasibility concerns and design choices conflict, they negotiate and compromise to protect potential RP impacts.

\section{User Reflections Stimulated by the Speculative Societies: Methods}
By engaging OSS designers and developers with the fictional societies we created, we aimed to understand how speculative design could affect practitioners' reflections related to OSS design. Through this process, we also wanted to gain deeper insights into ways of enhancing design contributions in OSS projects. The user study was approved by the research ethics board of all involved institutions.

\subsection{Participants}
We aimed to recruit participants with experience contributing to OSS within the past year as either a designer or a developer. We started by posting announcements in OSS communities, such as the Open Source Design forum, OSS-related subreddits, and LinkedIn. We later applied the snowball sampling method, asking active participants to refer others who fit the criteria. This helped us reach more participants with relevant experience. While we listed the required background in the ads and targeted emails, we further used a screening survey to ensure that the participants indeed fit our criteria. A total of 12 participants joined our study; this sample size was informed by thematic saturation during data analysis. Table~\ref{tab:study_participants} summarizes their characteristics. Participants' ages ranged from 18 to 60 years, with most between 26 and 45 years old. Their countries of origin covered Europe, North America, Asia, and Africa. Among the 12 participants, seven had experience involved in OSS as designers, while five as developers. Most of them ($N=8$) had more than three years of experience contributing to OSS.

\begin{table}[t]
\centering
\caption{Summary of characteristics of participants}
\resizebox{\textwidth}{!}{
\begin{tabular}{llllll}
\toprule
ID & Role & Years of OSS exp. & Types of OSS contributions & Age Grp. & Location \\
\midrule
P1 & Developer & 1 year or less & New features development & 18-25 & Canada \\
P2 & Developer & 2 to 3 years & New features development, documentation maintenance & 26-35 & Kenya \\
P3 & Designer & More than 3 years & UI/UX research and design, community lead & 36-45 & UK \\
P4 & Designer & More than 3 years & Design guide maintenance, community organization & 36-45 & Germany \\
P5 & Developer & More than 3 years & Maintenance of multiple projects, bug fixes & 26-35 & Netherlands \\
P6 & Developer & 2 to 3 years & Maintenance of self roject, translation & 18-25 & Iran \\
P7 & Developer & More than 3 years & Maintenance of multiple projects & 26-35 & India \\
P8 & Designer & More than 3 years & Design lead, UI/UX research and design & 36-45 & Germany \\
P9 & Designer & More than 3 years & UI/UX review, issue discussion, community organization & 36-45 & Germany \\
P10 & Designer & More than 3 years & Project lead, community lead & 36-45 & Germany \\
P11 & Designer & More than 3 years & UI/UX review and improvements, issue discussion & 46-60 & USA \\
P12 & Designer & 1 year or less & UI/UX review and improvements & 26-35 & Germany\\
\bottomrule
\end{tabular}
}
\label{tab:study_participants}
\end{table}

\subsection{User Study Procedure}
The user studies were conducted remotely using video conferencing to accommodate participants from different locations. Each session lasted around two hours. All study sessions were audio-recorded with the participants' consent and later transcribed for analysis.

At the beginning of the study, we conducted a short interview asking the participants about their roles, experiences, and values in OSS. We also asked them to discuss their existing thoughts on OSS design and current experience incorporating design and usability concerns in OSS. 

After the initial interview, each participant completed two sessions, with a ten-minute break between them. During each session, participants were introduced to one fictional society (\textit{Husia} or \textit{Reetar}). The order of the two societies was counterbalanced within each participant group (designers and developers) to mitigate fatigue effects. In each session, participants were first given a textual description of one of the fictional societies (see Appendix~\ref{sec:husia_description} and Appendix~\ref{sec:reetar_description}) and asked questions about how aspects of that society could positively or negatively impact OSS design and usability. They were then asked to reflect on their own OSS experiences and how their values aligned with those shown in the fictional society. Afterwards, participants watched a video storyboard (slides and video scripts are provided in Appendix~\ref{sec:husia_storyboard} and Appendix~\ref{sec:reetar_storyboard}) to help them engage more deeply with the lived experience of an individual in that society. We followed up with additional questions about their reflections on OSS contribution processes, the tools they use, and their general reflections regarding design involvement in OSS. After a ten-minute break, participants then engaged with the other society in the same manner.

\subsection{Data analysis}
We analyzed the user study data using a reflexive thematic analysis approach~\cite{thematicanalysis}. The analysis started with a round of familiarization with the data, through reading the transcripts, taking notes, and conducting discussions within the research team about our initial insights. We then conducted multiple rounds of coding collaboratively, led by the first author. During this process, we iteratively coded key ideas and insights from the data and discussed to refine and group those codes into categories. Towards the end of this process, we focused on further grouping the categories into broader themes, each of which contained several sub-themes and captured how the speculative societies stimulated participants' reflections on different aspects.

While we originally also wanted to capture participants' direct perceptions of the speculative design method, our analysis did not produce rich themes in this aspect. Although participants generally expressed interest and fascination toward speculative design, their direct reflections on it remained shallow. This might be due to the fact that this approach is mostly unfamiliar and novel to the participants, including the UX practitioners in our study who mostly worked within pragmatic, problem-solving–oriented design paradigms. Deeper, method-level reflections require longer exposure and familiarization, which could be initiated through the contributions of community pioneers and fostered with collective research and community efforts. Thus, our analysis focused on how the speculative societies facilitated participants' reflection on the subject matter that they are familiar with: OSS design.

\section{User Reflections Stimulated by the Speculative Societies: Results}
Our data analysis identified three main themes in participants' reflections stimulated by the speculative societies, including reflections on (1) the fundamental values regarding OSS contribution, (2) the barriers hindering the incorporation of design into OSS, and (3) the potential actions for improving OSS design integration.

\subsection{Reflections on the Fundamental Values Regarding OSS Contribution}

Through our fictional societies,  participants reflected on the values that guided their contributions to OSS, by either reassessing their prior beliefs or recognizing additional values that they had not considered or were not able to articulate before. 

\subsubsection{Opening the Process, Not Only the Product}
Before seeing our fictional societies, participants mostly considered transparency and openness as an innate characteristic of the distribution of OSS products. \textbf{After seeing the fictional societies, the importance of openness in collaboration and the design process became more apparent to the participants and was discussed by them as a key value.} For example, P7, after reflecting on Husia, said that: ``\textit{Everyone should feel welcome to contribute and modify and improve... And things should not change just because one [person] thinks about it, but [only after] some kind of discussion.}'' Some participants also reflected on how the openness value affected their own practice. For example, when reflecting on the open and universally accessible tools and knowledge available in Husia, P4 raised concerns about the dominance of proprietary design tools, noting that most designers rely on tools like Figma, which is ``\textit{not open source and has become more hostile to open source over time.}''

\subsubsection{The Intrinsic Value of Personal Growth}
Before seeing the speculative societies, some participants discussed the importance of personal growth through contributing to OSS, echoing the results in Section~\ref{subsubsec:designer_motivations}. \textbf{After seeing the fictional societies, some designers started to reflect on the intrinsic value of personal growth.} For example, P3, who did not initially mention this as their value, was inspired by Husia and shared their value about growth through exploration: ``\textit{Experimentation and research to find out what is needed happens in open source---that makes it a really important value.}'' Similarly, P10 reflected on Reetar's merit-based system in relation to their personal values and emphasized the importance of finding intrinsic values rather than relying on extrinsic measures of merit.

\subsubsection{Meritocracy is a Double-Edged Sword}
Before seeing the fictional societies, some participants saw the merit system as innate in OSS communities, as P10 summarized: ``\textit{You get merits basically by doing stuff, and by doing stuff people start to trust you.}''
\textbf{After seeing the speculative societies, participants critically considered meritocracy as a more complex concept.}
On the one hand, several participants emphasized the importance of individual recognition and expressed concerns that a lack of visible acknowledgement could reduce motivation to contribute. For example, reflecting on Husia where personal ownership does not exist, P5 was concerned: ``\textit{If I'm doing 100 hours of work in one week, I want to see the difference between myself and someone that's doing nothing.}'' On the other hand, participants acknowledged that an overemphasis on status and merit could lead to tensions within groups. For example, reflecting on Reetar, P9 initially mentioned their interest in a credit system that includes designers, but then pointed out that status-driven systems can negatively impact collaboration: ``\textit{Work will change if you want your contributions recognized... That's the standard issue with a metric that [...] starts becoming a goal in itself.}''

\subsubsection{A Broadened Scope of Community} Before exploring the speculative societies, participants primarily described the importance of community in terms of reaching out to users and connecting with other practitioners who have common interests. \textbf{After seeing our speculative designs, participants expanded their understanding of community, focusing more on inclusiveness and equity of the community members.} For example, P4, inspired by the fact that in Husia all creations are accessible and adaptable for diverse skill levels and needs, stressed the responsibility of contributors to consider less-represented users:  ``\textit{There is such an important thing to do that's like orienting yourself around people that can't fix that stuff themselves, but they depend on it.}''

\subsection{Reflections on the Barriers Hindering OSS Design Incorporation}
The alternative scenarios encouraged participants to have deeper insights into the root causes of the barriers in OSS design.

\subsubsection{Misconceptions Toward Designers and Their Work}
Evoked by the speculative scenarios, one key reflection from participants was that the presence of misconceptions about designers and their work within the OSS community contributes to the challenges designers encounter in OSS. They explored these misconceptions in detail and discussed factors that may have influenced how design and designers' work are perceived.
Participants discussed \textbf{a vicious cycle between designer involvement and their perceived status and value in OSS communities}. For example, after reflecting on Reetar, where developers cannot risk skipping design expertise to ensure product quality, P3 mentioned: ``\textit{A major concern in current open-source projects is that since designers don't touch the code, they won't stay engaged long-term. [...] This creates a cycle where designers aren't included in key discussions or treated as integral team members---leading them to leave, which in turn reinforces the original assumption.}'' 
Participants also noted that \textbf{one reason behind the perception that design work in OSS is costly and complex may be the late involvement of designers}. For example, P7, reflecting on Reetar, where designers are involved from the start of each project, elaborated on this point: ``\textit{When an OSS project finally approaches a designer [late in the process] and the designer suggests changes based on user research, it not only becomes a larger task but can also lead to communication breakdowns. Developers may feel like the designer is trying to change everything...}'' Moreover, participants critically reflected on how \textbf{the scope of design is often misunderstood in OSS}. They noted that OSS projects tend to reduce design work to superficial, visual-related contributions. For example, reflecting on Reetar, where designers play a central role in development, P2 observed: ``\textit{When I look at OSS repositories, I see feature flags for design, but actual design contributions are minimal. [...] Most design contributions I see are things like logos and badges.}''

\subsubsection{The Conceptual Paradox of OSS} Another notable barrier discussed by the participants was the inconsistency between what OSS communities claimed to value and the actual practice.
Participants reflected on \textbf{the illusion of open access for OSS users}, based on what they saw in the speculative societies. For example, reflecting on Reetar, where the platform and system support designers' contributions, P9 pointed out that while OSS claims to be open to all, its structure creates barriers for some: ``\textit{The code is available, but you need to compile it yourself, navigate forums, and have a specific type of operating system that it works on. So it's basically saying, `Only if you're that kind of person or become that kind of person---a hacker---then the software is accessible to you.'}'' 
The fictional societies also provoked participants' reflection on \textbf{the illusion of inclusiveness for OSS contributors}. While OSS communities often claim to be inclusive and welcoming, they sometimes restrict participation and reflect only certain contributors' views. For example, P1, a junior developer, shared their personal experience of starting in OSS without prior experience and reflected on Reetar's reputation points system: ``\textit{If someone with a low reputation [RP] successfully contributes and completes a task, will those with higher RP on the same project actually accept their patch? That's a major challenge in our current world. Maybe it's the same for designers, too.}''

\subsubsection{Information Barriers to Entry for Designers} 
After engaging with the speculative societies, participants discussed the challenges designers face when entering OSS and explored the possible reasons behind these difficulties. They explained in detail how \textbf{the complexities of contributing to an OSS project are often not apparent to new designers}, highlighting hidden factors such as project structure, workflows, and licensing complexities. For example, inspired by Husia's onboarding process, P1 explained their concerns towards designs in the current OSS landscape: ``\textit{As a junior developer, contributing to Linux feels overwhelming. Even small changes require deep project knowledge.}''
The speculative societies also prompted the participants to reflect further on how the \textbf{current OSS environment fragments access to resources related to design}. For example, reflecting on Reetar, where all decisions are documented in one place, P4 described the effort required to gather information in the current OSS: ``\textit{People have to dig through different channels, like Telegram or GitHub issues, and piece everything together on their own, which is a huge effort.}'' 
Moreover, engaging with the speculative societies allowed participants to reflect on why designers overlook OSS as a space to contribute, linking it to \textbf{cultural differences with traditional design practices}. For example, after reflecting on Reetar and how reputation functions there, P11 described:  ``\textit{Designers, instead of approaching [design in OSS] as a group, tend to focus on competing to have their work selected. The mindset of collaborating and sharing credit typically doesn't come at the beginning.}''

\subsection{Reflections on Actions to Take to Improve Design Integration in OSS}
As participants discussed challenges in OSS, they also reflected on their own practices, which led them to consider features from the speculative societies and imagine how similar ideas could benefit our current OSS ecosystem.

\subsubsection{Systematic Actions}
Participants highlighted actions needed at both structural and cultural levels of OSS, emphasizing the collective effort required to address OSS design challenges.

\textbf{Need for influential voices and resources:}
Participants emphasized that real progress to improve OSS design depends on those with power in the OSS ecosystem.
For example, P3, reflecting on Reetar, where success and recognition depend on reputation points and structured investment, highlighted: ``\textit{One impactful step would be for influential voices in open source, such as the Linux Foundation or major OSS organizations, to invest more in usability and design. ... We also need the backing of a big foundation or one of the major platforms like GitLab or GitHub. ... We need somebody with big RP!}''
Similarly, P5, reflecting on Husia, where resources and knowledge are shared, but long-term sustainability depends on collective contributions, mentioned: ``\textit{The companies that rely on OSS can support the projects by donating money. This funding could help projects hire more people, including maintainers and designers, leading to better software development.}''

\textbf{Open design collaboration:}
Another key aspect participants discussed was finding ways to enhance open collaboration in design.
For example, P4, reflecting on the centralized nature of tools in Husia, highlighted the need for more standardized design practices: ``\textit{Open standards could go much further. Imagine having a shared standard for UI design tools or illustrations that anyone could use!}''
Similarly, P1, reflecting on Reetar's Reputation system, pointed out:
``\textit{If someone has a low RP, I might judge them without realizing it---like how developers judge designers. A system with anonymous feedback could help reduce this bias and make evaluations more fair.}''

\textbf{Attributing credits to designers:}
Lastly, participants discussed the need for better ways to recognize designers' contributions in OSS. They criticized the lack of formal recognition for design contributions and imagined ways to acknowledge design work. For example, P1, as a developer, reflecting on Reetar's reputation system, stressed the importance of fairly acknowledging designers' work: ``\textit{I believe credit should be distributed fairly: the designer could receive about 65-70 percent of the reputation, while I take the remaining 30 percent, since the core idea came from them.}''

\subsubsection{Project-Level Actions} 
While reflecting on the fictional societies, participants discussed several ways in which OSS projects can make it easier for designers to contribute.

\textbf{Streamlining the onboarding process for designers:}
Participants discussed potential approaches to help designers get started in OSS projects. These approaches include creating better design onboarding materials and reaching out to potential contributors.
For example, P6, reflecting on Husia, where new contributors are seamlessly onboarded through the Hive's structured process, emphasized the importance of proactive engagement: ``\textit{We need to engage people who aren't typically involved in open-source projects. [...] We can do it in different ways, such as writing blog posts, sharing on social media, or creating videos about the contribution process.}''

\textbf{Reinforcing design-related task management:}
Participants also discussed various improvements in project management to better incorporate design into OSS projects. They advocated both clearer tracking for design tasks and stronger design leadership. For instance, reflecting on the way the Hive in Husia fosters collective work without strict individual ownership, P5 raised concerns about leadership and accountability: ``\textit{One issue I had with the Hive is that it's unclear who is in charge of each project... I believe it would be beneficial to have a designated individual for design---someone who can oversee the project's design, make key decisions, and resolve conflicts when needed.}''

\subsubsection{Tooling} Participants discussed specific features of tools they believed would enhance OSS design if incorporated. 
They pointed out that the current OSS process lacks \textbf{a centralized space for design documentation}. For example, reflecting on the platform Aplat, which allowed people in Reetar to document both conversations and design artifacts in one place, P10 mentioned: ``\textit{It would be a dream to have a tool like this in open source software, to keep everything together...}''
Participants also emphasized the importance of tools that enable \textbf{recording, synthesizing, and tracing design rationales} throughout the OSS process. For example, reflecting on Husia, where developers can assess design feasibility, project objectives, and user study results simultaneously, P3 mentioned: ``\textit{I've thought about what a tool might look like if developers, upon receiving the designs, could select parts of the UI and get automatically linked to the relevant research, documentation, or user needs that informed the decision[...]---to better integrate design rationale and research.}''
Moreover, participants evaluated various ideas to improve \textbf{design artifact integration} in OSS.
For example, P4 reflected on Reetar, where it was said that `people can build on each other's work.' They highlighted the potential of a system similar to GitHub for design, saying: 
``\textit{If we could have GitHub-style branching and merging for design files, it would be an incredible benefit. Imagine someone submits a contribution suggesting a small design change, and with a single click, it seamlessly integrates into the main design files...}''
Participants also emphasized the importance of \textbf{maintaining openness and embracing a diversity of tools} within OSS. For example, P8, reflecting on Husia, where design and development take place within a single, highly integrated system (the Hive and Xagon), expressed concerns about over-reliance on one tool, stating: ``\textit{I don't like the idea of one platform being used by all software practitioners... That gives too much control to the platform, and it could easily exploit its users.}''

\section{Discussion}
Our study showed that speculative design can be a powerful method to stimulate deep reflections on complex sociotechnical issues, such as the multifaceted roles of designers and their work in OSS. By drawing out people's underlying assumptions and future visions, this method creates spaces for reimagining possibilities for transformation~\cite{Auger}. In our study, speculative design not only encouraged participants to reflect on the processes and structures surrounding design in OSS, but also prompted them to consider more critical and nuanced sociocultural dynamics within these communities. Prompted by the speculative scenarios, our participants challenged the status quo of ``openness'' in OSS, a critique that echoes with some very recent studies on the phenomenon~\cite{Frluckaj2024}. They also proposed solutions covering multiple layers, ranging from revolutionary systemic shifts to specific project and tooling actions. Below, we first discuss our own reflections on using speculative design in a practical study context, i.e., promoting awareness and mindset change regarding OSS design integration. We then present some practical recommendations derived from our study aimed at helping practitioners better integrate designers and their work into the OSS process.

\subsection{Considerations on Using Speculative Design to Stimulate Sociotechnical Reflections}
The worlds of Husia and Reetar were deliberately designed to feel both familiar to OSS contributors and drastically different from our own, inviting participants to step outside their everyday workflows. Although some participants noted that valuing designers more closely matched their aspirations for OSS, the novelty of a designer-centric environment, combined with extremes like Reetar's rigid credit system and Husia's centralized design tooling, created surprising alternatives to the existing sociotechnical relationships of many OSS contributors, triggering reflection by magnifying and challenging the unconscious values and cultural assumptions~\cite{sengers}. The dual capacity of speculative design, i.e., to critique and to envision, positions it as a critical tool for both analysis and innovation in sociotechnical contexts, such as the OSS environment. 

\textbf{Building tension between ideals and realities.}
In our speculative design, we deliberately avoided creating overly idealized or bleak scenarios. Instead, we strove for a nuanced representation of the complexities related to designer involvement. With the practical context of OSS, we recognized that even the most well-intentioned solutions can have unintended consequences. Through our study, we came to understand that every collaboration model, tool, and incentive mechanism has its trade-offs. We aimed to convey this complexity to our audience. For example, the real-life OSS context lacks a reliable mechanism for designers to receive credit for their contributions. We addressed this in \textit{Reetar} with the reputation points. But instead of presenting it as a perfect solution to the credit attribution problem, we indicated its potential drawbacks in \textit{Reetar}, such as the risk of creating an overly competitive environment or leading to an overemphasis on credit negotiation rather than real collaboration. A directly opposing solution to this problem is to avoid credit attribution once and for all, which is presented in \textit{Husia}. But still, this approach could raise important questions about motivation, recognition of effort, and the potential further devaluation of contributions. The user study results indicated the effectiveness of our efforts, with rich and complex reflections among participants, e.g., on the topic of meritocracy of OSS. Similar effects were seen in other elements of our design, such as centralized tools and onboarding processes.

\textbf{Addressing all stakeholders for inclusion.}
Our speculative design deliberately targeted both designers and developers. For instance, while the \textit{Husia} storyboard featured a designer's experience, \textit{Reetar} centered around a developer as its main character, thereby offering a complementary perspective that encouraged empathy and mutual understanding. The descriptions of both societies were also crafted to avoid privileging one type of stakeholder over the other, although design is indicated to be valued. These considerations were informed by findings of our forum analysis that integrating designers into OSS communities is hindered by both developer-centric culture and designer mindset issues. During the user study, these design choices created an empathetic connection between designers and developers, the two primary groups of OSS stakeholders. Participants not only reflected on their own practices and experiences but also discussed their observations of others' workflows, often demonstrating an appreciation for the experiences and challenges faced by the other group. Prior studies have established that the roles of OSS contributors are often fluid and sometimes hidden~\cite{Cheng2019roles,Trinkenreich2020}. Such empathetic connections have the potential to foster a more inclusive and civil~\cite{Ferreira2021} environment within OSS communities, where different roles can achieve better mutual understandings and collaborate more effectively. This, in turn, can lead to more sustainable OSS communities that attract and retain a more diverse range of contributors.

\textbf{Inspiring solution-oriented thinking.}
We acknowledge that reflection alone, without considering potential actions, would not drive meaningful change. Achieving designer inclusion in OSS is a complex issue that may require a fundamental and gradual shift in the OSS culture and process. This realization led us to explore speculative design in this practical, sociotechnical context. By incorporating system probes into our speculative societies and scenarios, we aimed to encourage participants to think critically about potential sociotechnical factors and solutions. In both speculative societies, we introduced concrete tools that facilitated collaboration and supported the design process: in \textit{Husia}, the tool supports contributor onboarding, tracks tasks, and traces artifacts, while in \textit{Reetar}, the tool matches contributors, manages communication, and supports review. These tools sparked interesting discussions during the user studies. As participants engaged with the speculative worlds and the probing systems, they began to question their own beliefs and assumptions, generating innovative ``what-if'' strategies, such as debating the need for an integrated tool to address design process challenges and discussing the implications of credit attribution. These discussions ultimately led to various ideas and considerations for possible solutions.

Overall, our experience with speculative design shows its strong potential to critique existing practices and inspire innovative solutions in complex sociotechnical contexts like OSS design. These insights provide important guidance for future research to further refine speculative design approaches for improving inclusion and collaboration in OSS and other related domains.

\subsection{Challenges and a Road Map for Applying Speculative Design to Benefit OSS}
While some of our participants, especially those who had design backgrounds, were indeed familiar with the concept of speculative design, none had actually applied or practically engaged with it. This is largely because their work has primarily focused on pragmatic, problem-solving paradigms of development and design. As a result, during the user study, we observed that although our speculative scenarios elicited a wide range of critical reflections on designer involvement and the OSS process, participants' reflections on the speculative design method itself remained limited to fascination and surface-level recognition of its potential. This highlights a direct challenge of applying speculative design in the OSS context, since we cannot expect regular OSS community members who have not engaged with this method to initiate the efforts independently. To address this, we envision several possible ways to move forward to support and guide initiatives aimed at helping OSS communities better engage with and benefit from the speculative design approach.

\textbf{Increasing exposure through design showcases.}
Designers, researchers, and advocates familiar with both the OSS context and speculative design could create showcases that illustrate the possibilities of using the speculative design method in OSS. Our work serves as one such example. These efforts would not only demonstrate the method's potential but also spark curiosity and inspiration in OSS contributors towards this approach. Over time, more exposure can encourage speculative thinking within the OSS community and motivate more experimentation, imaginative exploration, and reflection.

\textbf{Fostering discussion in dedicated forums and workshops.}
Creating both online and in-person spaces where OSS contributors can discuss speculative design practices, share experiences, and reflect on scenarios can help deepen engagement with this method. Existing platforms that support the UI/UX design aspect of OSS, such as the Open Source Design forum\footnote{https://discourse.opensourcedesign.net} and their regular workshops held at FOSDEM, demonstrate the value of such spaces. These platforms could be leveraged, and similar spaces could be established to initiate discussions specifically around speculative design. These discussions can surface diverse perspectives, uncover new insights, and gradually build a community comfortable using speculative design for exploration and reflection.

\textbf{Creating toolkits to directly aid speculation in OSS.} Similar to design toolkits that provide prompts and structured activities to support speculative thinking in other contexts, such as the work by \citet{candy2018future} and \citet{Sadeghian2025}, speculation toolkits incorporating OSS values and considerations can be created. Our speculative design focused on the individualist-collectivist tension in OSS, which can be an element to be included in such a toolkit. By offering practical entry points, these toolkits can help contributors more easily engage with speculative practices for exploring alternative futures, reflecting on current structuration, and imagining new possibilities.

\subsection{Practical Recommendations for OSS Practitioners}
Participants' reflections in our study offer a series of practical recommendations that OSS practitioners can adopt to promote design incorporation. These recommendations can help overcome obstacles identified in our formative study and previous literature to improve collaboration, streamline workflows, and better integrate design into their projects.

\subsubsection{Recommendations for Designers}
Designers play a significant role in improving OSS design but may encounter unfamiliar practices and unwelcoming norms within OSS communities~\cite{ nichols2003, hellman_facilitating_2021}. Beyond learning more about OSS, the following strategies can be helpful:
\begin{itemize}
    \item \textit{Embrace the open design values}: By embracing the practice of sharing early design concepts, unfinished work, and improvements of others' work as part of an open design approach, designers can better align their practices with the open source values, thus better incorporating their work in the OSS process.
    \item \textit{Seek opportunities on issue tracking systems}: Issue tracking systems are an important part of every project, where different stakeholders raise concerns, discuss solutions, and manage tasks. Many participants mentioned that designers can often find ``\textit{clues}'' in these discussions. By looking at the problems people talk about, designers can see where their skills are needed.
    \item \textit{Showcase design work in developer-friendly formats}: Both designers and developers mentioned that annotated mockups that connect design elements to specific user needs or usability issues can guide developers. Short ``design walkthrough'' videos are also considered effective for explaining changes to maintainers and developers who may not be able to have live discussions. Such materials also make it easier to understand the rationales without needing to go through additional sources.
    \item \textit{Offer guidance and mentorship}: After accumulating some experience in OSS projects, designers could create guides to direct newcomers to design-friendly issues and provide suggestions on topics such as how to ask questions or manage feedback. Informal workshops and virtual meetups are additional ways to introduce new designers to the OSS culture and workflows.
\end{itemize}

\subsubsection{Recommendations for Developers} 
Developers play an essential role in encouraging design contributions and recognizing designers' efforts. Rather than sticking to the functionality-oriented mindset or only ``fixing'' design in the last minute~\cite{bach,nichols2003,wang}, developers can make an explicit effort to welcome designers and promote design contribution in their project:
\begin{itemize}
    \item \textit{Provide clear onboarding for designers}: Straightforward, step-by-step instructions for running the software would enable designers to test it and discover ways to contribute without facing complex setups. Labeling beginner-friendly tasks (e.g., ``\textit{good first design issue}'') could guide newcomers to feasible entry points.
    \item \textit{Make gestures to welcome designers}: Listing design-related tasks in the issue tracking system (e.g., ``\textit{UI mockup needed}'') and stating that non-code contributions are welcome in project documentation helps signal Inclusiveness. A dedicated ``Design Contributions'' section in the README or CONTRIBUTING file can also make this explicit.
    \item \textit{Collaborate proactively with design tools}: Creating a shared workspace (e.g., on Figma or Penpot) and inviting designers to offer feedback can keep design discussions organized and transparent. Consolidating constructive feedback within these platforms could centralize the rationale for design decisions and ensure a more transparent, traceable design process.
    
    \item \textit{Acknowledge and highlight design work}: Recognizing designers as core contributors when they significantly shape the interface or user experience can encourage further involvement. Public acknowledgments, through release notes, documentation, or social media, could reinforce the value of design. In commit messages, referencing user benefits or improvements would make design decisions more visible and respected.
    \item \textit{Promote the value of design}: Developers who value design should advocate its importance at OSS workshops, conferences, or online platforms. Presentations or discussions on design benefits can help shift perceptions and encourage greater support for design contributions.
\end{itemize}

\subsubsection{Structural Recommendations for the OSS Communities}
Moving beyond a fixation on code-centric workflows that often marginalize design work~\cite{bach, terry, hellman_facilitating_2021}, the broader OSS ecosystem can also adjust its infrastructure and culture to support design contributions more effectively:
\begin{itemize}
    \item \textit{Develop tools for collaborative design}: Introduce version control for design artifacts, where design assets can be proposed, branched, and merged like code. Enhance existing OSS platforms (e.g., GitHub or GitLab) so mockups and usability reports can be stored, reviewed, and tracked alongside code.
    \item \textit{Promote exemplars of successful OSS design}: Influential OSS organizations such as the Linux Foundation or Mozilla could sponsor design fellowships. These fellowships would encourage experienced designers to mentor others, create resources, and support OSS projects. Sharing success stories at OSS events can help normalize design as part of open source.
    \item \textit{Create inclusive merit systems}: Expand code-based reputation systems to acknowledge design activities (e.g., mockups and user tests). Provide special recognitions (such as ``\textit{UX Contributor}'') to highlight and reward design contributions.
\end{itemize}

\subsection{Limitations and Future Work}
Our study has revealed several promising directions for future research in integrating speculative design with OSS development practices. One critical avenue for exploration involves longitudinal studies to assess the lasting impact of speculative design interventions. While our current work provided valuable insights into immediate reactions, future research should examine how these interventions influence designers' engagement and collaboration over time. This extended view could help identify sustainable methods for integrating design into OSS workflows and better understand community dynamics. Since our study highlighted actions that can be systematically taken, another direction could be running participatory design sessions using these suggestions to envision their implementation and examine their challenges. Incorporating diverse perspectives is also crucial. Our participants are mostly from individualistic cultures themselves and have already had experience participating in OSS. Future research should include a broader range of participants, particularly those with more diverse cultural values and from underrepresented groups in OSS communities. Expanding the cultural and demographic scope of participants, combined with a larger sample size, could allow comparison of perspectives based on participants' own background and cultural tendencies. This would yield more comprehensive insights into the effects of speculative design, as well as systemic barriers and opportunities in OSS design, ultimately contributing to more equitable participation.

\section{Conclusion}
In this paper, we demonstrate how speculative design can surface hidden assumptions and foster reflections on the integration of design expertise within OSS communities. To inform the speculative design, we first analyzed forum posts to understand designers' core motivations and key challenges to participating in OSS communities. We then crafted two speculative societies, Husia (a collectivist society) and Reetar (an individualist society), to re-imagine how designer and their work can be valued and supported. User studies with 12 OSS practitioners (seven designers and five developers) with these scenarios suggest that our speculative scenarios prompted participants to re-examine fundamental OSS values and identify the root causes behind existing barriers. Through reflective discussions, practitioners proposed a range of future actions and recommendations. Together, our work not only yielded concrete guidelines for fostering more designer-inclusive OSS environments but also provided insights into the use of speculative design in practical sociotechnical contexts.

\begin{acks}
    We thank our participants for their time and valuable insights. We also thank the anonymous reviewers for helping us improve the paper. This work is supported by the Alfred P. Sloan Foundation (G-2021-16745), the Canada Research Chairs program (CRC-2021-00076), and the Natural Sciences and Engineering Research Council of Canada (NSERC).
\end{acks}

\bibliographystyle{ACM-Reference-Format}
\bibliography{references}

\appendix
\newpage
\section{Society Description and Storyboard for Husia}
\label{sec:husia_details}
\subsection{Husia's Society Description}
\label{sec:husia_description}
Husia is a small, interconnected society where community and collaboration are central. The people of Husia know each other by name and face. Personal ownership and monopolization are not common concepts. Everything is shared by all. Work in Husia is done in groups, and ideas come from working together instead of from one person. The main goal is to create things that are useful, not to find out who came up with the idea.

Children in Husia attend the same school from an early age, receiving a standard education, with professional concentrations that take into account their individual traits and abilities. Each child is guided toward a profession suited to their strengths, with a common focus on contributing to society in their chosen field.

Openness is fundamental in Husia. All products and knowledge are freely available for anyone to use, modify, and improve. This approach extends to the software development processes, where all creations are made available to everyone, ensuring they can be used, adapted, and understood by individuals of all skill levels and needs.

Design plays an important role in Husia to create products that help everyone, especially in the software development process. The Hive, a building dedicated to technological development in Husia, is where people with expertise in design, development, and other fields work together. The Hive is powered by an intelligent system and equipped with interactive smart devices that assist professionals with their tasks. The building is organized into separate spaces with smart walls known as the Xagon. The Xagon stores, organizes, and manages data, tasks, and project results. Each person working in the Hive has a specific profile connected to them, which is updated based on their activities and their previous background in education. The system learns and adjusts to each worker's style and needs while maximizing the expected benefit of the entire society, providing specific recommendations to optimize the workflow.

\subsection{Husia's Storyboard}
\label{sec:husia_storyboard}
\begin{figure}[h!]
    \centering
    \includegraphics[width=0.8\linewidth]{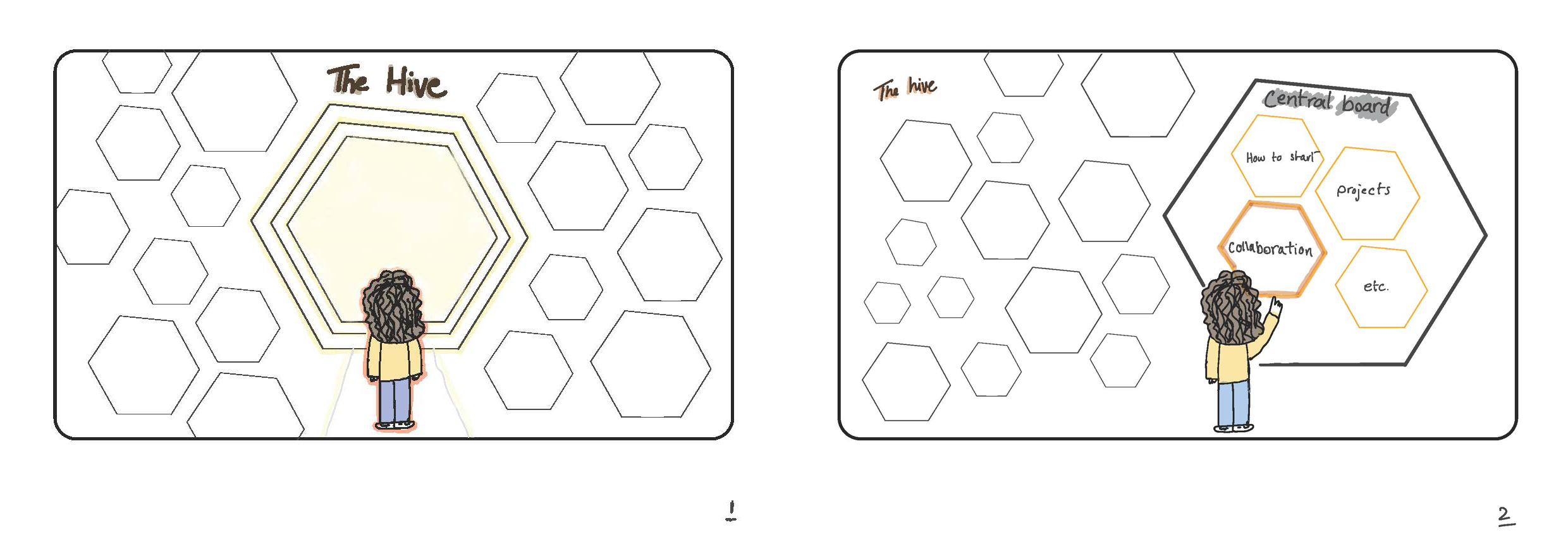}
    \includegraphics[width=0.8\linewidth]{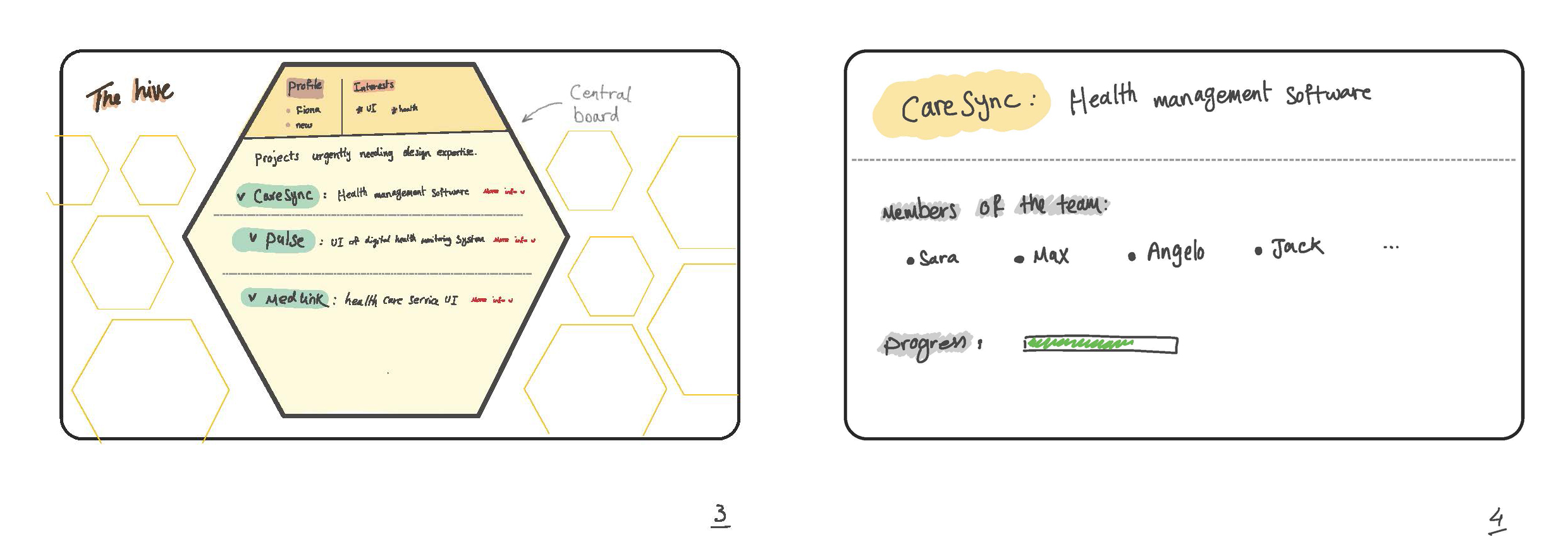}
    \includegraphics[width=0.8\linewidth]{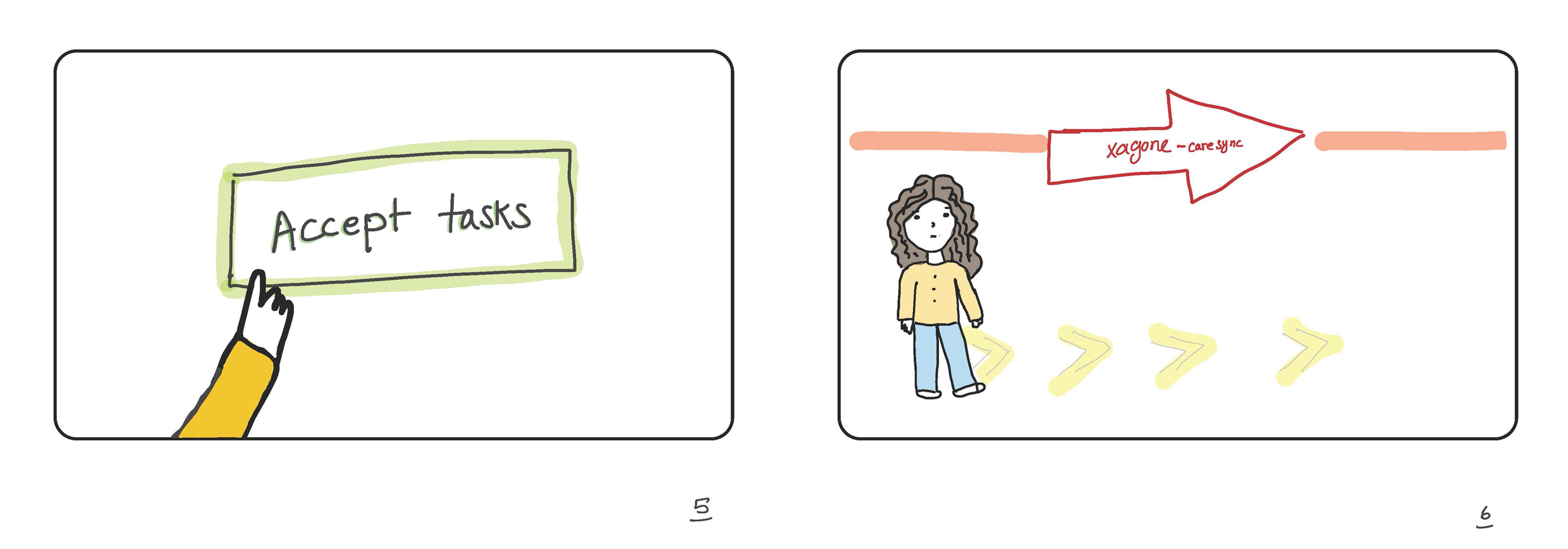}
    \includegraphics[width=0.8\linewidth]{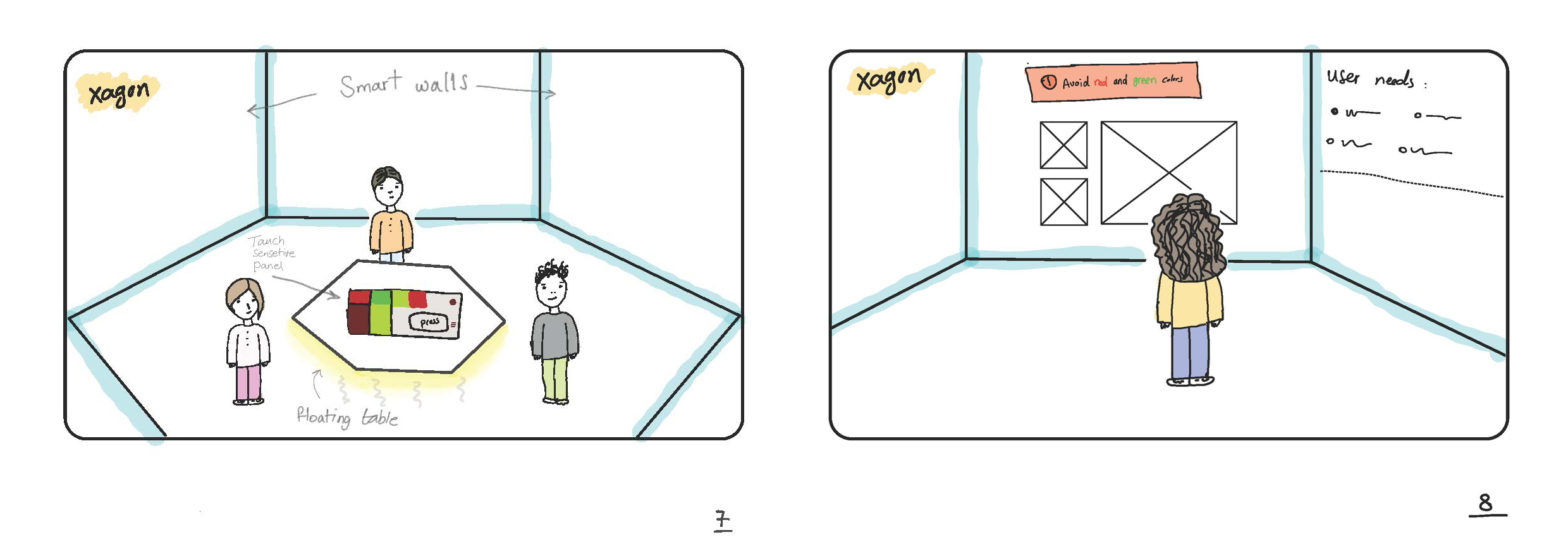}
    \caption{Husia's Storyboard}
    \label{fig:husia_storyboard_1}
\end{figure}

% \newpage
\begin{figure}[h!]
    \centering
    \includegraphics[width=0.8\linewidth]{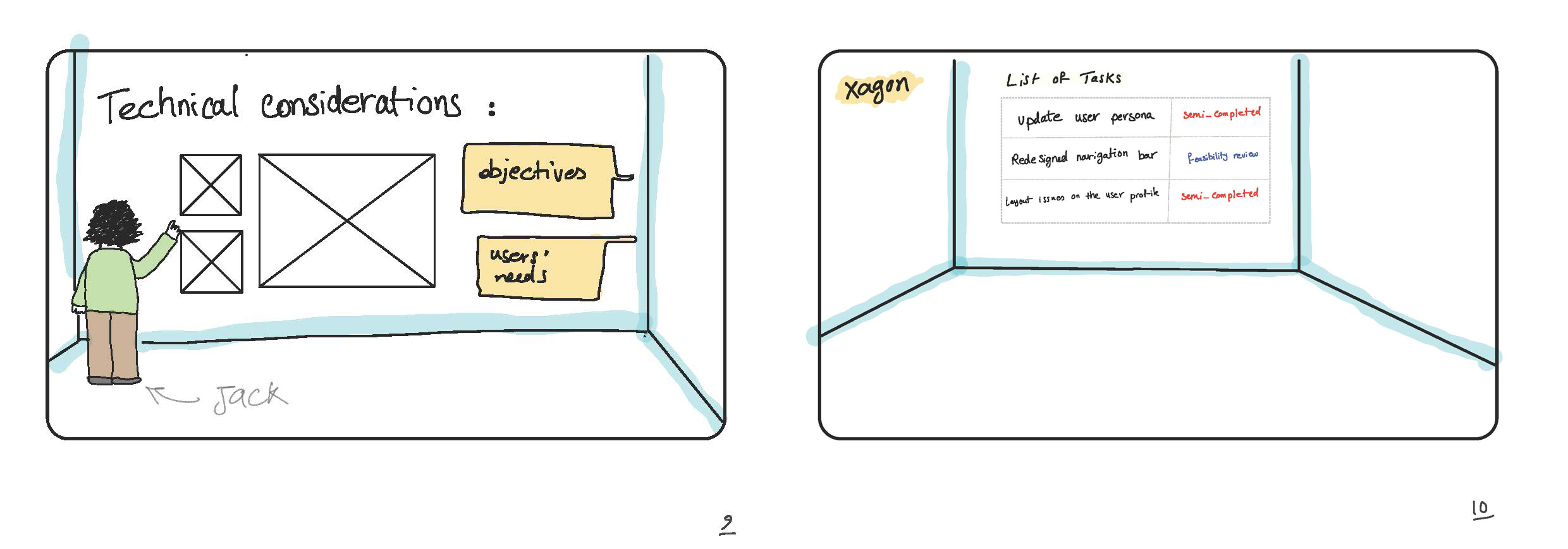}
    \caption{Husia's Storyboard (continued)}
    \label{fig:husia_storyboard_2}
\end{figure}

The Husia storyboard visuals are shown in Figures~\ref{fig:husia_storyboard_1} and~\ref{fig:husia_storyboard_2}. Below is the script used in a recorded slideshow video to present the storyboard, with the visuals as slides.

Frame 1. Fiona, a recent graduate from design school, walks into the Hive. Upon entering, the Hive identifies her as a new person and guides her to the Central Board.

Frame 2. The interactive screen there explains how to start working on a project, how to collaborate, and how to join a project. Anything she needs to start contributing to design-related tasks is provided by the central board.

Frame 3. Since Fiona is a first-time contributor, the Central Board shows her the projects in the Hive that need immediate help with design expertise, including those in health and education.

Frame 4. After the Central Board provides more details about each project, including its size, current progress, and other members of the team, Fiona chooses a concrete health management software project to join, given her strong interest in this domain and her belief that she can work seamlessly with other team members.

Frame 5. Then the Central Board matches the specific tasks in graphic and UI design in the project to Fiona's profile in the system and prompts those tasks for Fiona to confirm and accept. 

Frame 6. After Fiona's confirmation of the project and tasks, the Central Board directs Fiona to the Xagon associated with that project. There, she joins the group to work on this project right away.

Frame 7. Xagons are rooms in the Hive, each of which is dedicated to a project. They act as smart support systems specifically for each project, where team members physically work together. In addition to providing collaboration spaces, Xagons track versions of both design and code, helping new team members like Fiona understand the project's history and progress. Using Xagon's real-time features, team members working on design and code regularly provide feedback to each other. Additionally, Xagon helps team members collect and summarize user feedback and ensure the team remains informed about user needs.

Frame 8. Upon reviewing the list of tasks, Fiona selects a task to address a few usability issues on the landing page, as it currently poses important challenges for some users. Fiona works on new layout options to improve usability and also changes the color scheme and fonts based on user feedback. The Xagon helps by reminding her about user needs and the main types of users for this product. For example, it warns her to avoid using red and green to indicate button meanings, as 67 percent of reported users are colorblind and have trouble differentiating between these colors. She also follows the system's advice to check the project's goals and design rules, which were set by the previous team members and saved in Xagon.

Frame 9. Once the redesign draft is ready, Xagon marks it as a coding task so people with relevant expertise can see it in their task list. The person working on the code reviews the design to check if it is technically feasible and identifies any possible challenges with implementation, such as performance. Throughout the process, Xagon reminds the person working on the code of the design objectives and their impact on the users, helping them understand the importance of the design decisions made by Fiona.

Frame 10. After completing the redesign, the Xagon marks it as semi-completed, bringing it to the attention of other team members for review and comments.

\section{Society Description and Storyboard for Reetar}
\label{sec:reetar_details}
\subsection{Reetar's Society Description}
\label{sec:reetar_description}
In Reetar, reputation is the currency. Reputation is measured for each person by RPs, which are determined by the reactions of others to the service or product one offers. RPs determine individuals' social status and the opportunities they have. People are born with different RP depending on their parents' status. 

The stakes are high. While people could gain RPs by offering good services or products, a single mistake can reduce RPs permanently and affect one's standing in society. If they work alone, the impact on their RP will be solely on the individual, but if they work in a group, the impact is distributed among the team. As a result, while some still decide to go solo to maximize their RP gain, especially when they have a unique idea and the competence to realize it, many people in Reetar tend to work in groups to mitigate the risk. 

To earn a significant amount of RP, the project must be innovative, as simply replicating existing products or services results in lower rewards. At the same time, improving or building on other people's work is also common, especially when building on a product that has already made a positive impact. Both the original creator and the new contributors can earn RP if the new product performs well.

Aplat is a platform used by all software practitioners in Reetar that connects developers, designers, and other experts to work together. Contributors earn RP when a project is successful and receives positive feedback, but lose RP otherwise. Designers hold an important position because they ensure a product, especially software, meets user expectations, and thus receives positive reactions from the users. Since any success and failure can impact RP, people are cautious and selective about the projects they join and the people they work with. As a result, while Aplat enables collaboration, it remains a competitive environment. Developers and designers must balance innovation with the risk of failure.

\subsection{Reetar's Storyboard}
\label{sec:reetar_storyboard}
\begin{figure}[th]
    \centering
    \includegraphics[width=0.8\linewidth]{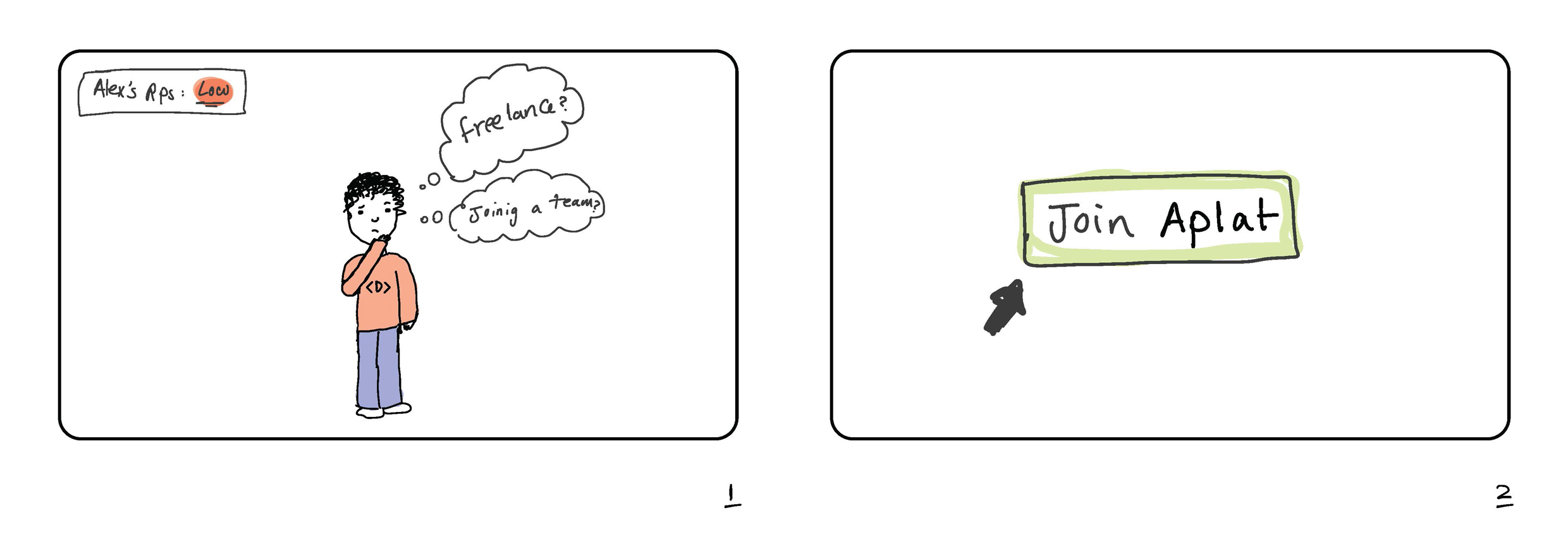}
    \includegraphics[width=0.8\linewidth]{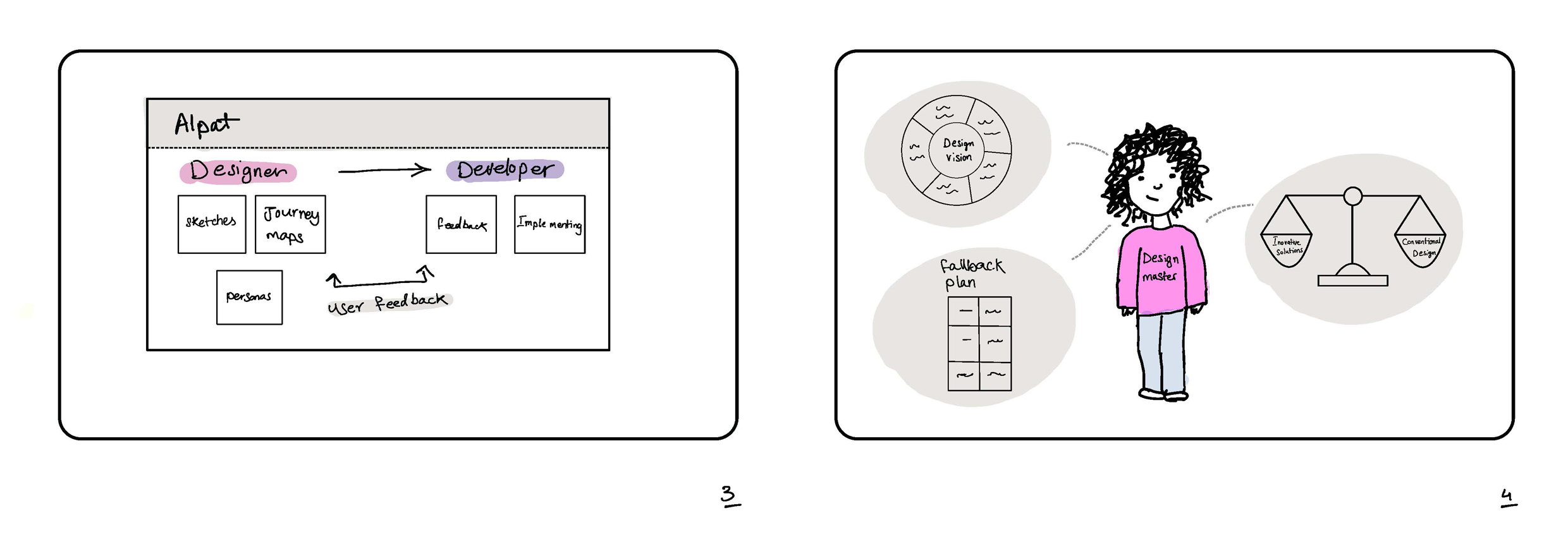}
    \includegraphics[width=0.8\linewidth]{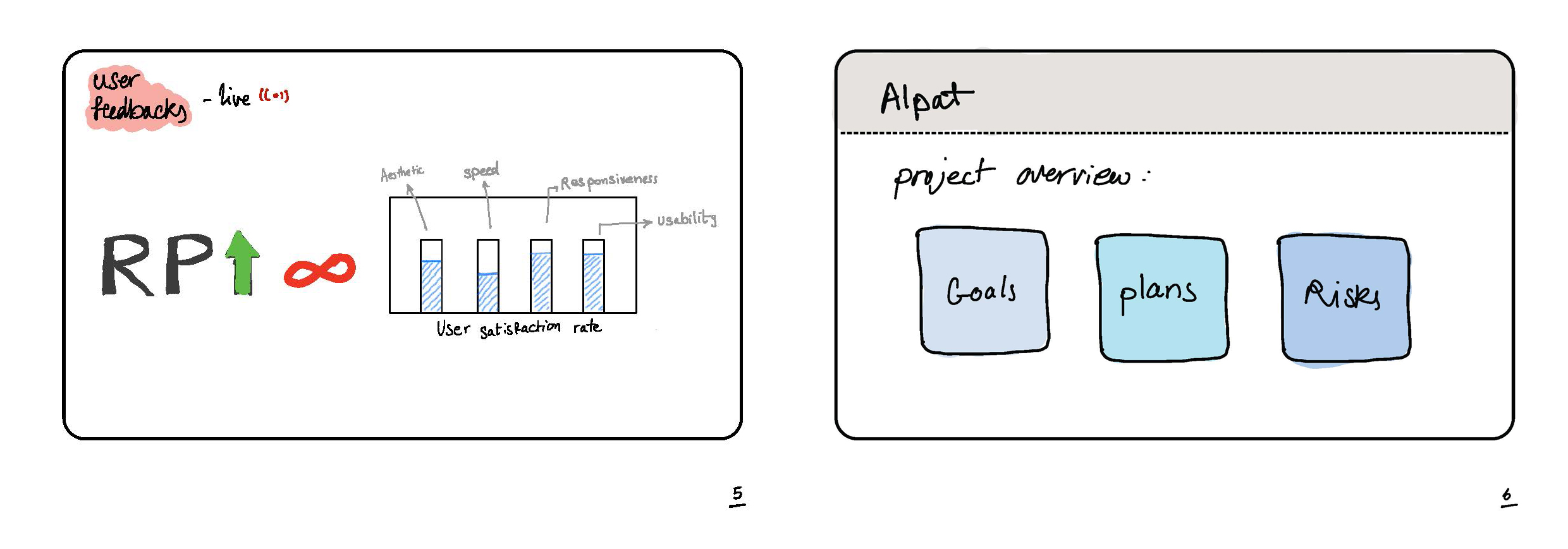}   \includegraphics[width=0.8\linewidth]{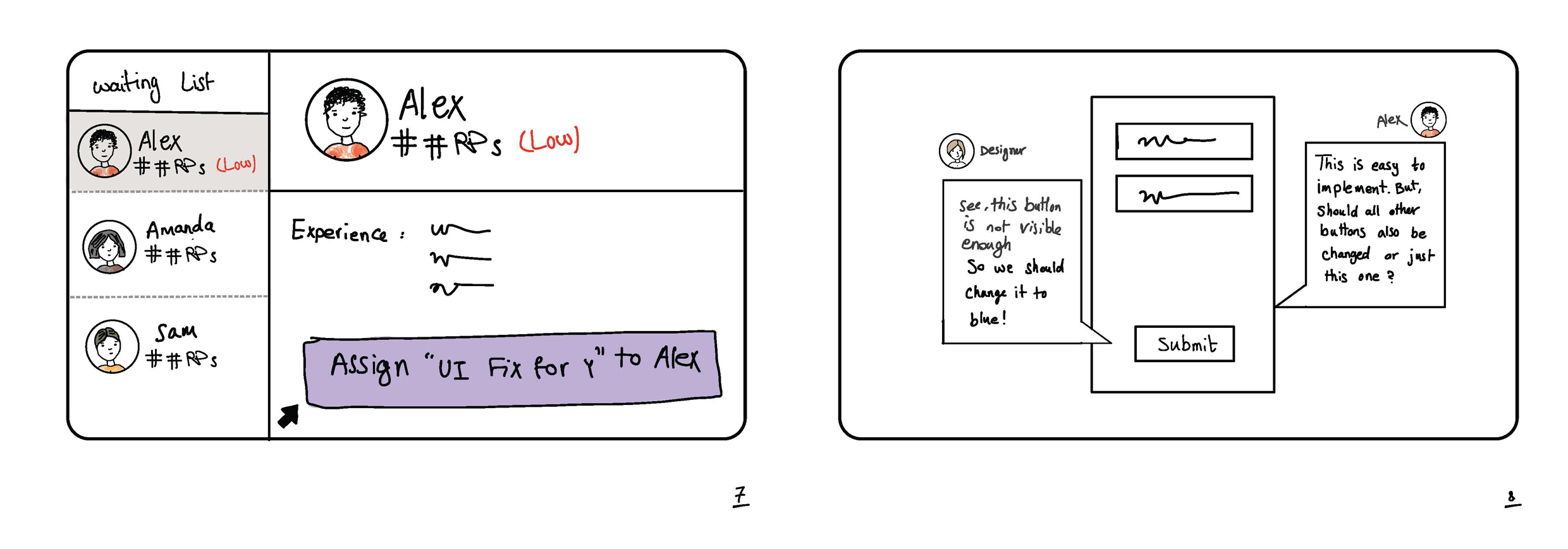}
    \caption{Reetar's Storyboard}
    \label{fig:reetar_storyboard_1}
\end{figure}

% \newpage
\begin{figure}[ht]
    \centering 
    \includegraphics[width=0.8\linewidth]{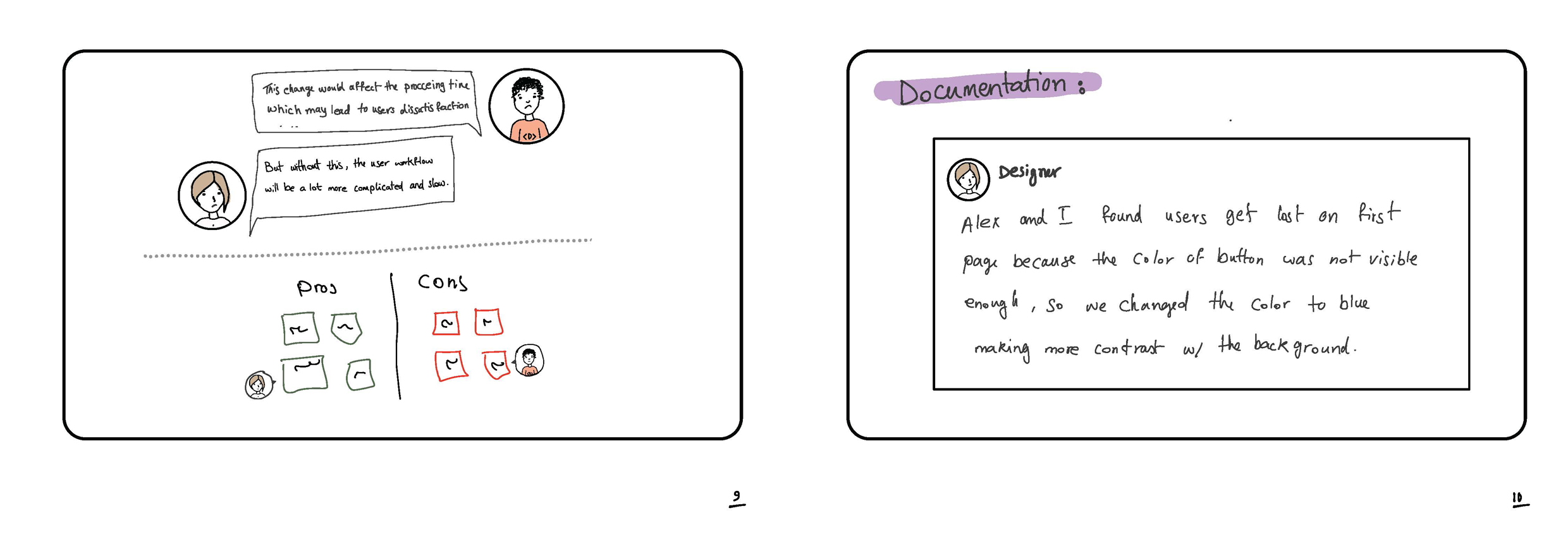}
    \caption{Reetar's Storyboard (continued)}
    \label{fig:reetar_storyboard_2}
\end{figure}

The Reetar storyboard visuals are shown in Figures~\ref{fig:reetar_storyboard_1} and~\ref{fig:reetar_storyboard_2}. Below is the script used in a recorded slideshow video to present the storyboard, with the visuals as slides.

Frame 1. Alex is a recent graduate and wants to become a developer. To start his career, he has two options: freelance or work with a team. 

Frame 2. Freelancing involves higher risks, as there is great uncertainty in gaining user approval for his software. So, he decides to increase his RPs by joining a team project. He chooses a project he is knowledgeable about using Aplat. 

Frame 3. Aplat is structured around the design process, with built-in features that emphasize user feedback and usability. For instance, the development process could not start without sketches and other artifacts created by the designers. Aplat also incorporates detailed user feedback loops, requiring developers and designers to collect real-time insights into user interactions with the software. Furthermore, Aplat includes built-in tools such as user persona creators and journey mapping features, enabling teams to better understand their audience. 

Frame 4. Every project includes a design master, who is responsible for balancing user needs with RP goals. Also sets the design vision and ensures every decision aligns with RP priorities, while preparing fallback plans to avoid penalties.

Frame 5. RP directly impacts every step of the design and development process. Designers and developers carefully evaluate decisions based on user feedback and RP risks. For example, features that could result in user dissatisfaction are adjusted to protect the RPs. User testing is critical and stressful, as it has a direct impact on RP outcomes.

Frame 6. Aplat records all project goals, plans, and risks, ensuring the team stays aligned and new members can evaluate a project's RP risks before joining. Every update, from wireframes to final maintenance, prioritizes user feedback while minimizing RP loss. 

Frame 7. Back to Alex. After indicating his interest in joining a team, he is put on a waiting list to be matched with designer-initiated tasks. For some, this process could last for months. Luckily for Alex, a designer with a similar low level of RP joined the team, created designs to solve some problems, and is waiting for a developer to implement them.

Frame 8. They start working on the task using a space that Aplat provides, an integrated environment for design and development where they can collaborate and communicate simultaneously. Designers can see in real-time how the UI is being implemented and provide feedback, while developers can comment on the design offered by the designers. 

Frame 9. Naturally, issues and disagreements arise along the way. Sometimes, Alex worries that the designer's ideas are not feasible and could negatively impact their RPs, so he requests changes and provides explanations. However, the designer insists that the feature they proposed is necessary based on user feedback. In such cases, they engage in rounds of discussion, with both Alex and the designer compromising and exploring different ways to keep the solution feasible while aligning with user preferences and RPs.

Frame 10. The entire communication and collaboration process is documented so that other developers and designers can reference the reasoning behind the solution. Once they are finished, they post their solution in a section for tasks pending review. Other designers and developers check the solution, and if it receives enough approval, it is added to the original software.

\end{document}